\documentclass{article}

\usepackage{arxiv}

\usepackage[utf8]{inputenc} 
\usepackage[T1]{fontenc}    
\usepackage{hyperref}       
\usepackage{url}            
\usepackage{booktabs}       
\usepackage{amsfonts}       
\usepackage{nicefrac}       
\usepackage{microtype}      
\usepackage{lipsum}
\usepackage{graphicx}
\usepackage{placeins}
\usepackage{amsmath}
\usepackage{algorithm}
\usepackage{algpseudocode}
\graphicspath{ {./Figures/} }
\usepackage{natbib}
\usepackage{pifont}
\usepackage[dvipsnames]{xcolor}
\setcitestyle{authoryear,open={(},close={)}} 

\usepackage[shortlabels]{enumitem}

\title{Machine Learning-Guided Design of Non-Reciprocal and Asymmetric Elastic Chiral Metamaterials}

\author{
 Lingxiao Yuan \\
  Department of Mechanical Engineering\\
  Boston University\\
  \texttt{lxyuan@bu.edu} \\
   \And
 Emma Lejeune
 \footnotemark[1]\\
  Department of Mechanical Engineering\\
  Boston University\\
  \texttt{elejeune@bu.edu} \\
    \And
Harold S. Park
 \thanks{corresponding authors.} \\
  Department of Mechanical Engineering\\
  Boston University\\
  \texttt{parkhs@bu.edu} \\
}

\begin{document}
\maketitle
\begin{abstract}
There has been significant recent interest in the mechanics community to design structures that can either violate reciprocity, or exhibit elastic asymmetry or odd elasticity.  While these properties are highly desirable to enable mechanical metamaterials to exhibit novel wave propagation phenomena, it remains an open question as to how to design passive structures that exhibit \emph{both} significant non-reciprocity and elastic asymmetry.  In this paper, we first define several design spaces for chiral metamaterials leveraging specific design parameters, including the ligament contact angles, the ligament shape, and circle radius.  Having defined the design spaces, we then leverage machine learning approaches, and specifically Bayesian optimization, to determine optimally performing designs within each design space satisfying maximal non-reciprocity or stiffness asymmetry.  Finally, we perform multi-objective optimization by determining the Pareto optimum and find chiral metamaterials that simultaneously exhibit high non-reciprocity and stiffness asymmetry.  Our analysis of the underlying mechanisms reveals that chiral metamaterials that can display multiple different contact states under loading in different directions are able to simultaneously exhibit both high non-reciprocity and stiffness asymmetry.  Overall, this work demonstrates the effectiveness of employing ML to bring insights to a novel domain with limited prior information, and more generally will pave the way for metamaterials with unique properties and functionality in directing and guiding mechanical wave energy.
\end{abstract}


\section{Introduction}
There has been significant recent interest in the topics of reciprocity~\citep{nassarNRM2020}, and elastic asymmetry within the broader mechanics community.  Reciprocity implies that if we push a structure on one side (X), the other side (Y) will move by a certain amount.  If we push the opposite side (Y) with the same force, side X will move the same amount.  This idea has been codified through the well-known Maxwell-Betti reciprocity, which can be written as
\begin{equation}\label{eq:mb} F_{X}u_{Y\rightarrow X}=F_{Y}u_{X\rightarrow Y}
\end{equation} 
There is significant interest in finding structures that can break reciprocity, either statically~\citep{coulaisNATURE2017,shaatSR2020}, or dynamically~\citep{wangNC2023,nassarJMPS2017,trainitiNJP2016,wangPRL2018,goldsberryJASA2019,fangPRA2021,luEML2021,luEML2022,patilEML2022,wallenSAV2018,kuznetsovaNPG2017,attarPRA2020}.  The major motivation for this has been related to non-reciprocal wave propagation, in which structures enable wave propagation in one direction, but support different levels of wave propagation in other directions.  Most approaches to accomplishing this have involved active structures, in which the elastic stiffness and/or density can be actively modulated in space and time to enable non-reciprocal wave propagation~\citep{nassarJMPS2017,trainitiNJP2016,wangPRL2018}.  In contrast, passive approaches to generating non-reciprocal structures have been less studied, due to challenges in creating passive structures that violate reciprocity~\citep{wangNC2023,goldsberryJASA2019,fangPRA2021,luEML2022,attarPRA2020}.  This has generally been accomplished using passive structures that exhibit a nonlinearity in which the elastic stiffness of a structure is bilinear, or different depending on the direction of loading~\citep{wangNC2023,luEML2022,goldsberryJASA2019}.

In addition to non-reciprocal elasticity, there has recently emerged significant interest in creating structures that exhibit asymmetric elasticity~\citep{scheibnerNP2020,chenNC2021,tanNATURE2022,yinJE2023,zhangJMPS2020}.  This interest has emerged because the mechanical behavior of linear elastic, isotropic solids is typically described by a free energy function, which carries the implication that its elasticity tensor is symmetric.  However, Scheibner et al. recently proposed the notion of odd elasticity, for those linear elastic isotropic solids whose mechanical behavior cannot be described by a free energy function~\citep{scheibnerNP2020}. As a result, odd elastic solids have a non-symmetric elasticity tensor, where the mechanical response to different loads (in contrast to the difference in mechanical response to different directions for non-reciprocal elasticity) is not the same.  For example, in such a solid extension could induce shear, while the same shear would induce a different amount of extension.  It has been shown that such odd elastic solids could induce interesting dynamic phenomena, including non-Hermitian wave propagation~\citep{chenNC2021}.  However, similar to non-reciprocal solids, it has been considerably easier to achieve asymmetric elasticity using active, rather than passive solids.

Recently,~\cite{shaatJMPS2023} proposed a chiral metamaterial which exhibits both non-reciprocal and asymmetric elasticity.  While chiral metamaterials have been widely studied over the past decade~\citep{wuMD2019,corbatonAM2019,liuJME2016,liuJMPS2012,chenIJSS2013,nassarPRL2020,shaatJMPS2023}, the mechanism enabling this behavior is that of contact, in which the ligament connecting two rigid circles is initially in a state of contact with both circles.  Non-reciprocal (i.e. directional) elasticity is realized because while the ligament remains in contact with both circles under tension, and thus is stiff, it loses contact with the circles under compression, and is thus elastically soft.  Similarly, asymmetric elasticity results because in certain directions of loading, the ligament remains in contact, while in other directions of loading it does not.  This, in conjunction with the chirality that couples different deformation modes, enables asymmetric elasticity to occur.  

An important open question that remains is whether we can create passive solids with tunable mechanical properties that are both non-reciprocal and asymmetric, and further what the mechanisms are that would allow such a combination of properties to be realized.  Because the design space for the chiral metamaterial is relatively large, encompassing potential factors such as circle diameter, ligament contact area, ligament geometry, one approach to realizing these properties is by utilizing a machine learning model to learn the combination of factors enabling these properties to be realized using starting with the base two circles, one ligament system.  

In this paper, we first define several design spaces for chiral metamaterials leveraging specific design parameters, including the ligament contact angles, the ligament shape, and circle radius.  Having defined the design spaces, we then leverage machine learning approaches, and specifically Bayesian optimization, to determine optimally performing designs within each design space satisfying maximal non-reciprocity or stiffness asymmetry.  Finally, we perform multi-objective optimization by determining the Pareto optimum and find chiral metamaterials that simultaneously exhibit high non-reciprocity and stiffness asymmetry.  Our analysis of the underlying mechanisms reveals that chiral metamaterials that can display multiple different contact states under loading in different directions are able to simultaneously exhibit both high non-reciprocity and stiffness asymmetry.  Overall, this work demonstrates the effectiveness of employing ML to bring insights to a novel domain with limited prior information, and more generally will pave the way for metamaterials with unique properties and functionality in directing and guiding mechanical wave energy.

\section{Chiral Metamaterial}

\begin{figure}[h!]
    \centering
    \includegraphics[width=.5\textwidth]{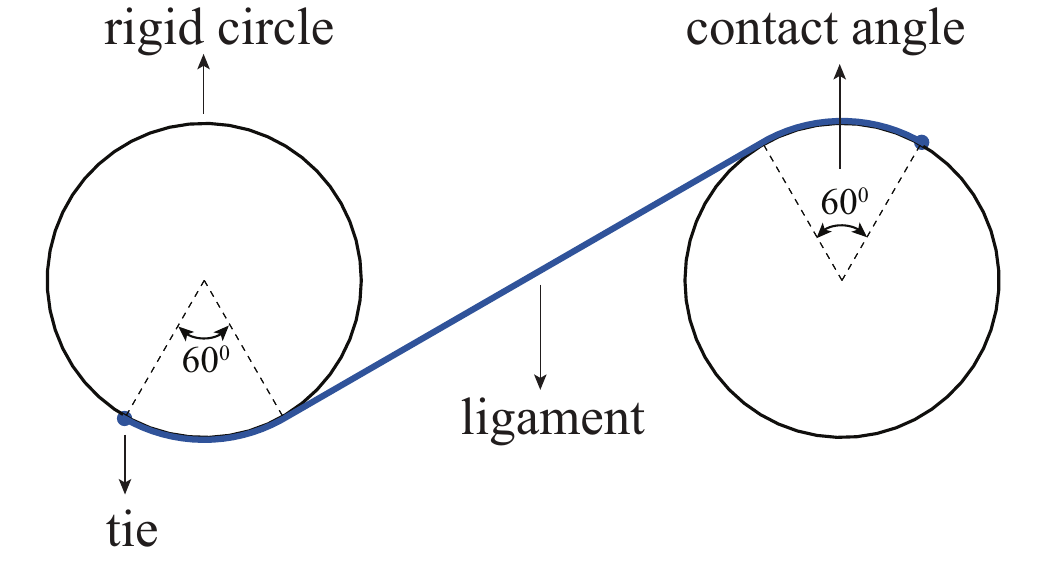}
    \caption{Illustration of the chiral metamaterial in ~\cite{shaatJMPS2023}. The ligament was tied to two rigid circles at the ends. The ligament shape was fixed and the contact angle was $60^0$ at both sides.} 
    \label{fig:origin}
\end{figure}

~\cite{shaatJMPS2023} introduced a chiral metamaterial that exhibits nonreciprocal and asymmetric elasticity due to changes in the internal contact mechanism when forces are applied from different directions. The fundamental unit of this chiral metamaterial consists of two rigid circles connected by an elastic ligament. The surface of the rigid circle is frictionless. In ~\cite{shaatJMPS2023}, this building block is characterized by a straight ligament that has a fixed contact angle with the rigid circles, as shown in Fig. \ref{fig:origin}. Note that the initial contact is stress-free so there is no contact pressure between the ligament and the circle. In this paper, we will show that the response to force from different directions is significantly influenced by the contact between the ligament and the rigid circles. To harness the potential of the ligament-circle contact mechanism and facilitate programmability in the material's reciprocity and stiffness asymmetry, we expand the chiral metamaterial design space by introducing additional variations in ligament shape, contact angle, and circle radius. Here, we lay out the details of this mechanical system. 

\subsection{Stiffness Definition}
\label{sec:stiffness}
\begin{figure}[h!]
    \centering
    \includegraphics[width=.9\textwidth]{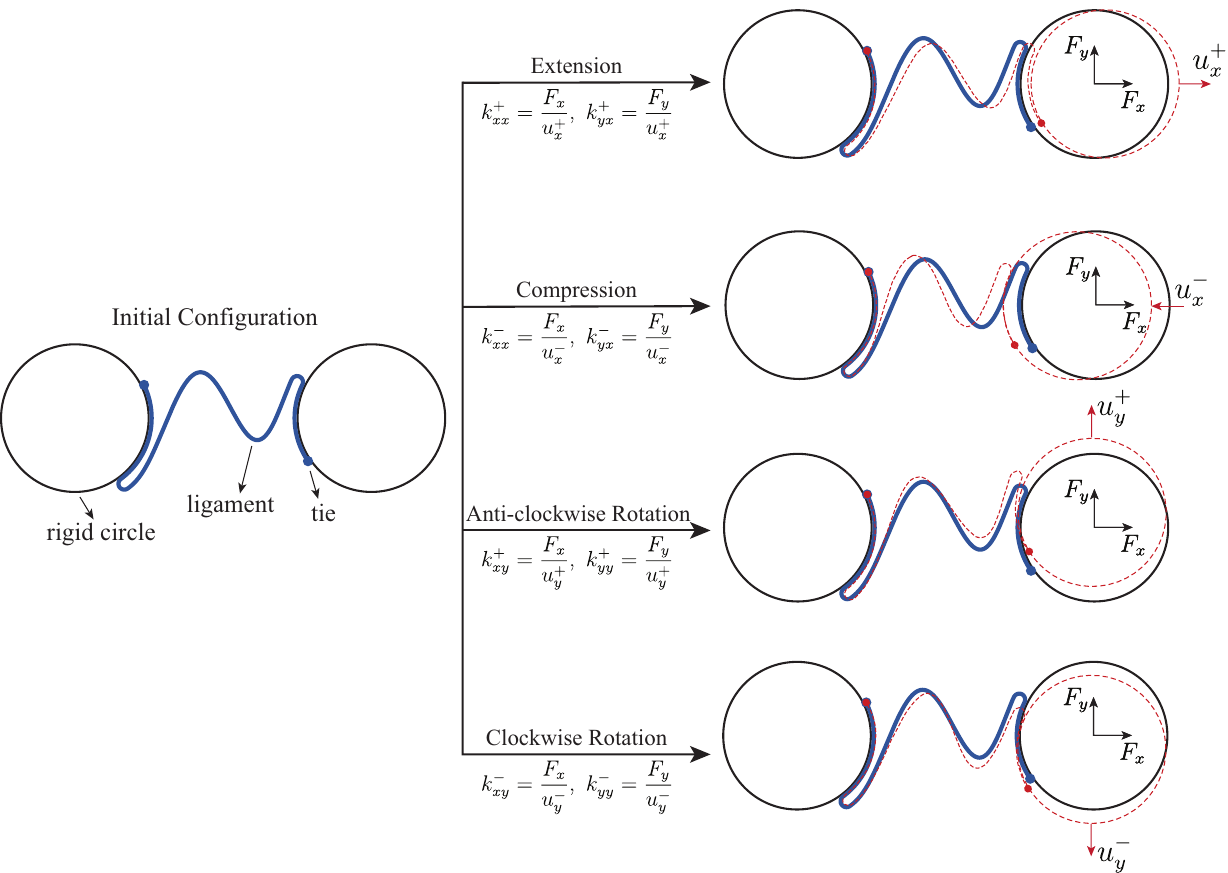}
    \caption{Illustration of the chiral metamaterial undergoing displacement from four directions. The reaction force of the right rigid circles in the $x$ and $y$ directions are denoted as $F_x$ and $F_y$. The stiffness values are calculated using the formula $k_{ij} = F_i/u_j$ when $u_j$ is nonzero. A superscript ``$+$'' is added to the displacement and stiffness symbol when the displacement is positive, and ``$-$'' when the displacement is negative.} 
    \label{fig:stiff}
\end{figure}

Fig. \ref{fig:stiff} illustrates a representative structure undergoing four different types of deformation: extension, compression, anti-clockwise rotation, and clockwise rotation. Finite Element Analysis (FEA) ~\citep{liu2021eighty,szabo2021finite,yang2022rate,yuan2020effects,yuan2019notification} using the commercial software ABAQUS ~\citep{smith2009abaqus} was conducted to obtain the force-displacement response for each deformation (see \ref{apx:fea} for more details). For extension, a displacement $u_x^+$ to the positive $x$ direction was applied to the right rigid circle. From reaction forces computed via FEA, we can get two stiffness values $k_{xx}^+ = F_x/u_x^+$, where $F_x$ is the reaction force in the $x$ direction when the displacement is in the $x$ direction, and $k_{yx}^+ = F_y/u_x^+$, where $F_y$ is the reaction force in the $y$ direction when the displacement is in the $x$ direction. For compression, we apply a displacement $u_x^-$ in the negative $x$ direction. Similarly, we compute the stiffness values $k_{xx}^- = F_x/u_x^-$ and $k_{yx}^- = F_y/u_x^-$. To distinguish the stiffness values obtained from the deformation of different directions, we use the symbol ``$+$'' when the displacement is in the positive direction and the symbol ``$-$'' when the displacement is in the negative direction. For the anti-clockwise rotation, the displacement $u_y^+$ was applied and two stiffness values $k_{xy}^+ = F_x/u_y^+$ and $k_{yy}^+ = F_y/u_y^+$ are calculated. For the clockwise rotation, the displacement $u_y^-$ was applied and the two stiffness values $k_{xy}^- = F_x/u_y^-$ and $k_{yy}^- = F_y/u_y^-$ are calculated. Considering all of these displacement directions, we summarize the eight stiffness values of the chiral unit using an array $K = [k_{xx}^-,k_{xy}^-,k_{yx}^-,k_{yy}^-, k_{xx}^+,k_{xy}^+,k_{yx}^+,k_{yy}^+]$. We describe the material using the stiffness matrix:

\begin{equation}
\label{eq:sf_matrix}
    K = 
        \begin{bmatrix}
        k_{xx} & k_{xy}\\
        k_{yx} & k_{yy}
        \end{bmatrix}
\end{equation}

where the values of each element $k_{ij}$ can be either $k_{ij}^-$ or $k_{ij}^+$. For typical materials, the relationship $k_{ij}^- = k_{ij}^+$  and $k_{ij} = k_{ji}$ holds due to the Maxwell–Betti theorem ~\citep{viesca2011elastic, coulaisNATURE2017}. When the stiffness values are different in opposite directions of the same axis, i.e. $k_{ij}^- \neq k_{ij}^+$, the reciprocity of linearity breaks and the material behaves like a bilinear spring ~\citep{luEML2022}. When the stiffness matrix is asymmetric, i.e. $k_{ij} \neq k_{ji}$,  the reciprocity of the stiffness matrix is broken, similar to the odd elasticity described recently~\citep{scheibnerNP2020}. Both scenarios lead to nonreciprocal effects in wave propagation. To distinguish between these cases, we refer to $k_{ij}^- \neq k_{ij}^+$ as ``non-reciprocity'' and  $k_{ij} \neq k_{ji}$ as ``asymmetry''. In this work, we aim to design a chiral metamaterial to break symmetry and reciprocity to the largest extent, e.g., $k_{ij}^- >> k_{ij}^+$ or $k_{ij} >> k_{ji}$.

\subsection{Contact Mechanism}
\label{sec:contact}

\begin{figure}[h!]
    \centering
    \includegraphics[width=.9\textwidth]{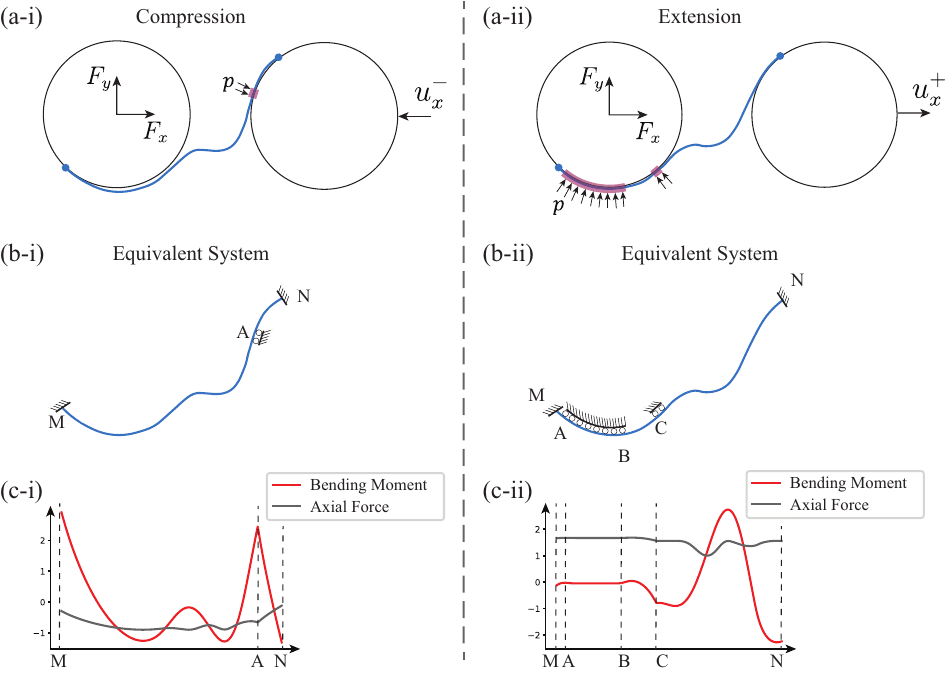}
    \caption{Mechanical model of the chiral structure under compression and extension. (a) Chiral structure under (i) compression load and (ii) extension load. The contact area of the deformed structure is highlighted with red color. (b) Equivalent mechanical model illustrating the contact mechanism with roller supporters substituting the contact area. (c) Distribution of Bending Moment and Axial Force along the beam.
}
    \label{fig:cont}
\end{figure}

In this Section, we investigate how loads applied from various directions can induce distinct contact modes between the rigid circle and the elastic ligament. An analogy can be drawn between these diverse contact modes and varying the boundary conditions of an elastic beam where adding and removing support conditions can dramatically change the mechanical state of the structure, resulting in asymmetrical properties under different loading modes. To further elucidate this concept, Fig. \ref{fig:cont} presents an example of an identical structure subjected to forces from two opposing directions -- compressive and extensional loading. In Fig. \ref{fig:cont}, ligament deformation was obtained through finite element simulation and magnified by a factor of 10 to aid in visualization.

Fig. \ref{fig:cont}(a-i) shows the deformed structure under compressive load. The left circle remains fixed while the right circle is subject to a displacement load $u_x^-$ towards the left. Following the deformation, the left circle and the ligament are detached while the right circle and the ligament have a small area of contact. As the circle is frictionless, the ligament in the contact area only experiences pressure from the perpendicular direction. Consequently, the mechanics of the structure can be equivalent to a beam fixed at both ends with a frictionless roller in the middle, as illustrated in Fig. \ref{fig:cont}(b-i). In this context, the bending moment around the roller support does not change since roller supports do not contribute to bending moments. Specifically, if a roller support is positioned at the midpoint of a beam, the bending moment will attain its maximum value and undergo an abrupt transition. To validate this, Fig. \ref{fig:cont}(c-i) shows the evolution of the bending moment and axial force along the beam from the left end to the right end. Notably, the bending moment around the roller support (point A) reaches a maximum, aligning well with the proposed model in Fig. \ref{fig:cont}(b-i) that describes the mechanism of the chiral structure. 

Fig. \ref{fig:cont}(a-ii) shows the deformed structure under an extension load. The left circle is fixed and the right circle is subject to a displacement load $u_x^+$ towards the right. Following the deformation, the right circle and the ligament are detached while the left circle and the ligament establish two contact areas. Similarly, the Fig. \ref{fig:cont}(b-ii) presents an equivalent model of Fig. \ref{fig:cont}(a-ii). The two contact areas, one from point A to point B, and another around point C, are also equivalent to roller supports. Fig. \ref{fig:cont}(c-ii) shows the evolution of the bending moment and axial force along the beam for the extension load. The bending moment remains constant in the area of rollers. 

While the ligament is linear elastic, variations in boundary conditions occur under loads from different directions, leading to varying stiffnesses in different directions. The goal of this paper is to maximize these discrepancies to achieve extreme mechanical behaviors. In the following Section, we will elaborate on the details of these objectives.

\subsection{Strain Energy Components}
In the previous Section \ref{sec:contact}, we provided a qualitative analysis of the contact modes of the chiral structures. Specifically, the non-reciprocity and asymmetry of chiral metamaterials are produced by the variation in contact modes leading to different boundary conditions of the equivalent beam models. In this Section, we conduct a quantitative examination of the stiffness disparities. Specifically, we delve into the mechanics of the deformation under different loading directions to understand how stiffness is either enhanced or weakened depending on the direction of applied loads.

\begin{figure}[h!]
    \centering
    \includegraphics[width=.5\textwidth]{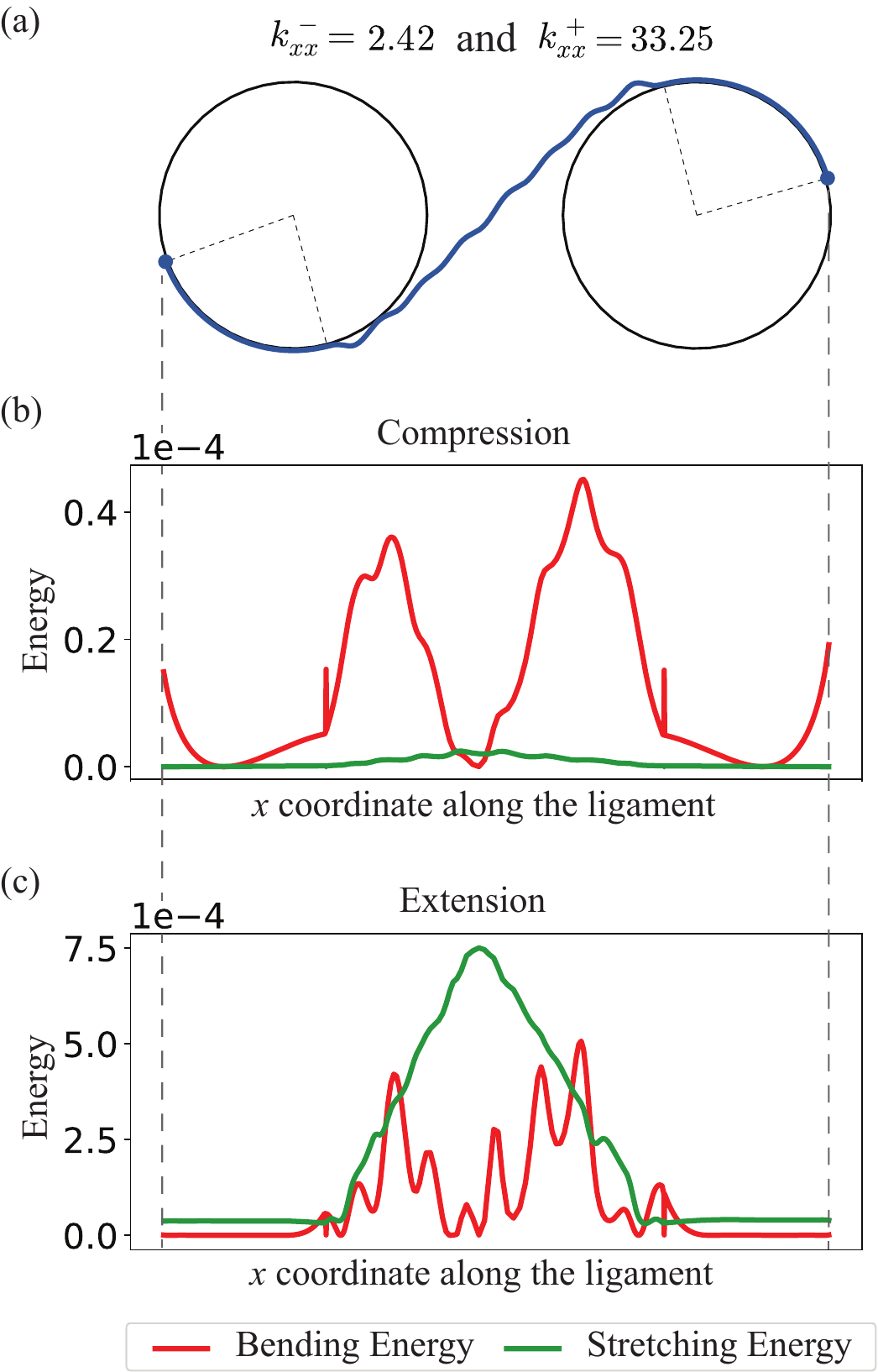}
    \caption{Bending and Stretching Energy distribution along the elastic ligament. (a) The schematic of the chiral structure. The design has a smaller stiffness $k_{xx}^-$ under compression loads, and a larger stiffness $k_{xx}^+$ under extension loads. (b) The bending and stretching distribution along the ligament under compression load. (c) The bending and stretching distribution along the ligament under extension load. The three figures (a)(b)(c) share the same \textit{x} coordinates.}
    \label{fig:energy}
\end{figure}

Fig. \ref{fig:energy}(a) shows a representative design where the value of $k_{xx}^-$ and $k_{xx}^+$ are different. Regardless of the directions, when the only nonzero displacement is $u_x$, the stiffness $k_{xx}$ of the structure is determined by evaluating the total change in strain energy by the equation below

\begin{equation}\label{eqn:stiff_strain}
    E_{strain} = \frac{1}{2}k_{xx}u_x^2
\end{equation}

Following the finite element analysis of this structure, the strain energy is $0.007764$ after compression and $0.1064$ after extension, with the $u_x$ set constant as $-0.08$ for compression and  $+0.08$ for extension. Through Eq. \ref{eqn:stiff_strain}, we derive the stiffness $k_{xx}^- = 2.42$ for compression and $k_{xx}^+ = 33.25$ for extension. It's evident that the strain energy during extension significantly surpasses that of compression, resulting in markedly higher stiffness during extension. Considering that bending and stretching energies are the primary components of strain energy for a linear elastic beam, we investigate the stiffness disparities by analyzing the stretching and bending energy distribution along the elastic ligament in Fig. \ref{fig:energy}(b)(c). During compression, the stretching energy is negligible compared to the bending energy, whereas during extension, the stretching energy dominates over the bending energy. This disparity indicates that ligament deformation is predominantly bending during compression, whereas stretching dominates during extension. Given the inherent difficulty in stretching a beam compared to bending it, the stretching energy is generally much larger than the bending energy, leading to a substantially greater resistance to external forces during extension. As a result, the stiffness value in extension is much larger than that of compression for this structure. By manipulating the geometry of the chiral structures, we can control the stretching and bending behavior under different loads, allowing us to discover optimal chiral structure designs capable of exhibiting extreme non-reciprocity and asymmetry.

\subsection{Optimization Objectives}
\label{sec:obj} 
As briefly introduced in Section \ref{sec:stiffness}, this work aims to enhance the nonreciprocity and asymmetry of the chiral metamaterial. This section provides additional details on the formalization of these objectives. Furthermore, the scope of the investigation will be extended to multi-objective optimization,  allowing for the identification of optimal designs that simultaneously exhibit both nonreciprocity and elastic asymmetry. 

\subsubsection{Non-Reciprocity}
\label{sec:obj_nonreci}

Following the definition in Section \ref{sec:stiffness}, enhancing the non-reciprocity of metamaterials equals maximizing the variance of stiffness values along opposite directions, i.e. $k_{ij}^- \neq k_{ij}^+$. To achieve this, we aim to maximize the ratio between the absolute values of the stiffness. The optimization task includes eight objectives to be maximized, outlined below:

\ding{172} $f_1 = |\frac{k_{xx}^-}{k_{xx}^+}|$, 
\ding{173} $f_2 = |\frac{k_{xx}^+}{k_{xx}^-}|$, 
\ding{174} $f_3 = |\frac{k_{xy}^-}{k_{xy}^+}|$, 
\ding{175} $f_4 = |\frac{k_{xy}^+}{k_{xy}^-}|$, 
\ding{176} $f_5 = |\frac{k_{yx}^-}{k_{yx}^+}|$, 
\ding{177} $f_6 = |\frac{k_{yx}^+}{k_{yx}^-}|$, 
\ding{178} $f_7 = |\frac{k_{yy}^-}{k_{yy}^+}|$, 
\ding{179} $f_8 = |\frac{k_{yy}^+}{k_{yy}^-}|$.

Note that when the value of these objectives is equal to $1$, i.e. the stiffness in opposite directions is the same, the corresponding structure exhibits reciprocity. Additionally, for each pair of stiffness $k_{ij}^-$ and $k_{ij}^+$ that we aim to maximize the difference between, we can either design a structure that $k_{ij}^-$ is larger $k_{ij}^+$, or $k_{ij}^+$ is larger $k_{ij}^-$. For instance, to maximize the disparity between $k_{xx}^-$ and $k_{xx}^+$, objective $f_1$ aims to identify an optimal design where $k_{xx}^-$ significantly exceeds $k_{xx}^+$, whereas objective $f_2$ seeks an optimal design where $k_{xx}^+$ significantly exceeds $k_{xx}^-$. Consequently, when searching for optimal designs that maximize non-reciprocity objective $f_1$ to $f_8$, only designs with objective values greater than $1$ are considered. In summary, our goal can be formalized as below:  

\begin{equation}
    \max f_i, ~\textrm{where} ~f_i > 1 ~\textrm{and}  ~i \in \{1,2,\ldots,8\} 
\end{equation}

\subsubsection{Elastic Asymmetry}
\label{sec:obj_asym}

Following the definition in Section \ref{sec:stiffness}, the asymmetry of chiral metamaterial is characterized by an asymmetric stiffness matrix, i.e. $k_{ij} \neq k_{ji}$. Similarly to the approach outlined in Section \ref{sec:obj_nonreci}, 
 we aim to maximize the ratio between the stiffness values. The eight objectives for asymmetry optimization are summarized below:  
 
\ding{172} $g_1 = |\frac{k_{xy}^-}{k_{yx}^-}|$, 
\ding{173} $g_2 = |\frac{k_{xy}^-}{k_{yx}^+}|$, 
\ding{174} $g_3 = |\frac{k_{xy}^+}{k_{yx}^-}|$, 
\ding{175} $g_4 = |\frac{k_{xy}^+}{k_{yx}^+}|$, 
\ding{176} $g_5 = |\frac{k_{yx}^-}{k_{xy}^-}|$, 
\ding{178} $g_6 = |\frac{k_{yx}^+}{k_{xy}^-}|$, 
\ding{177} $g_7 = |\frac{k_{yx}^-}{k_{xy}^+}|$, 
\ding{179} $g_8 = |\frac{k_{yx}^+}{k_{xy}^+}|$.

A more general form of these objectives is as below:

\begin{equation}
    \max g_i, ~\textrm{where} ~g_i > 1 ~\textrm{and}  ~i \in \{1,2,\ldots,8\} 
\end{equation}

\subsubsection{Multi-Objective}
\label{sec:obj_multi}
In Sections \ref{sec:obj_nonreci} and \ref{sec:obj_asym}, a single objective was employed to assess the performance of the chiral structures. However, optimizing a single objective guarantees design optimality in only one dimension. For multiple competing objectives, maximizing a single dimension may compromise the performance concerning another objective. To discover a chiral metamaterial that exhibits multiple novel functionalities, we strive to optimize the material to exhibit both non-reciprocity and elastic asymmetry. 

\begin{figure}[h!]
    \centering
    \includegraphics[width=.4\textwidth]{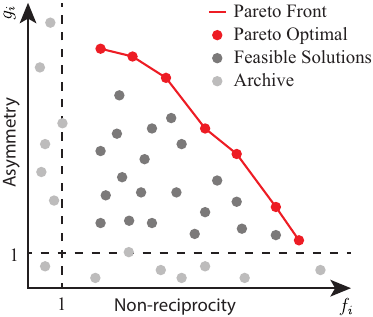}
    \caption{Illustration of Pareto front for the multi-objective optimization of chiral metamaterial.}
    \label{fig:pareto}
\end{figure}

The problem of searching for optimal designs that excel in multi-objectives is broadly known as determining the Pareto front ~\cite{schulz2018interactive, riquelme2015performance,li2024computational}. The design solutions to the multi-objective optimization are referred to as Pareto optimal. Specifically, a Pareto optimal design is one for which any adjustment cannot improve all the objectives simultaneously. In other words, for any Pareto optimal design, there is no alternative design that surpasses it in all performance functions. The set containing all the Pareto optimal designs is the Pareto set. The corresponding performance of the Pareto set in a performance space is the Pareto front. To facilitate understanding these concepts in the context of our problem, Fig. \ref{fig:pareto} illustrates a schematic performance map of design solutions. Each data point on the map is randomly generated and depicts performances $f_i$ and $g_i$ of a design solution, where $f_i$, representing an objective of non-reciprocity in Section \ref{sec:obj_nonreci}, and asymmetry $g_i$, representing an objective of asymmetry in Section \ref{sec:obj_asym}. In our problem setting, the objective values of the optimal designs must be larger than $1$. Consequently, the designs not satisfying this constraint will not be considered and are denoted as 'Archive'. The design solutions colored red are the Pareto optimal, as no other feasible solutions can achieve larger values for both $f_i$ and $g_i$. All other feasible solutions are colored grey. Our goal is to identify the Pareto optimal designs and the corresponding Pareto front to enhance both non-reciprocity and asymmetry. We expressed this goal mathematically as below:

\begin{equation}\label{eq:multiobj1}
    \max \{f_i, g_j\}, ~\textrm{where} ~f_i, g_j > 1, ~i,j \in \{1,2,\ldots,8\} 
\end{equation}

\subsection{Design Spaces}
\label{sec:ds}

\begin{figure}[h!]
    \centering
    \includegraphics[width=.9\textwidth]{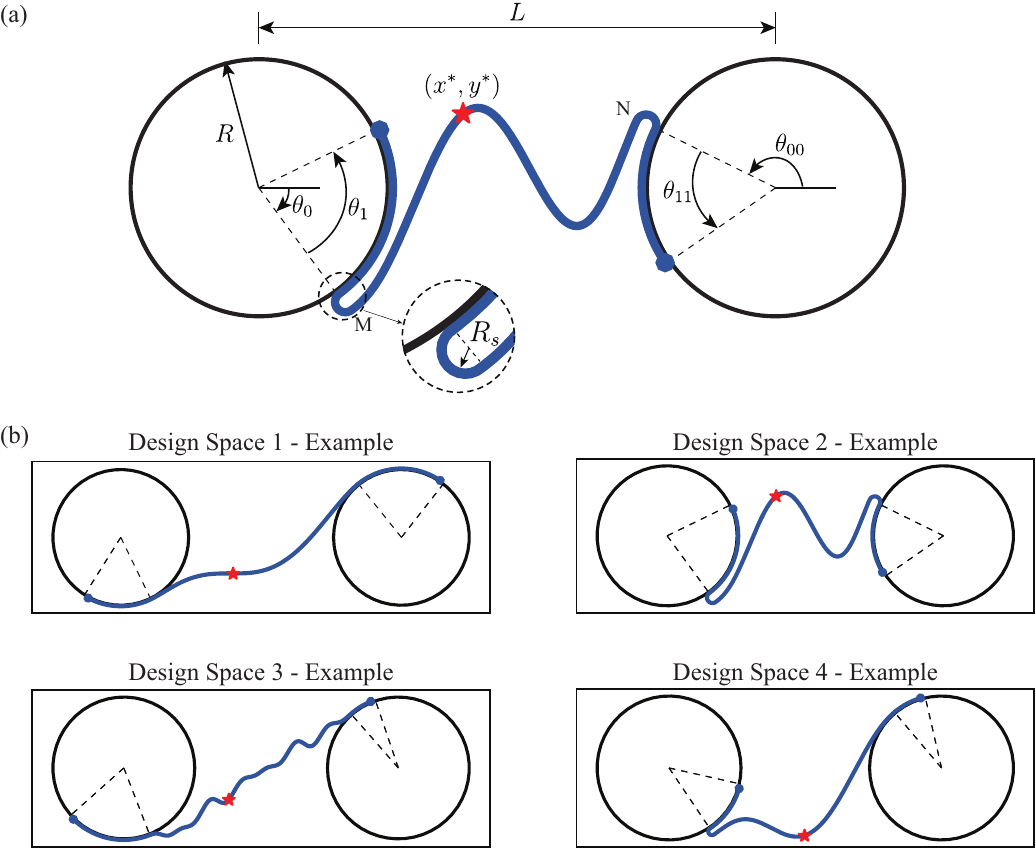}
    \caption{Illustration of (a) the design parameters and (b) the design spaces of the chiral metamaterial.}
    \label{fig:ds}
\end{figure}

As illustrated in Fig. \ref{fig:ds}(a), the programmable chiral structure consists of two rigid circles with the same radius $R$. The distance between the two circles is $L$, which has a constant value $20$. An elastic ligament initiates its connection with the left circle at an angle $\theta_0$ and forms a contact angle $\theta_1$. Similarly, the ligament establishes a connection with the right circle from an angle $\theta_{00}$ and forms a contact angle $\theta_{11}$. Positive values for the contact angle indicate an anti-clockwise orientation from the connecting point and vice versa for negative values. For the part of the ligament that is not in contact with the circle, there is a half circle with radius $R_s$ at the transition point for both sides. The ending point for the small half circle on the ligament is denoted as points M and N. The middle part from M to N is continuous and each point is sampled from the following function:

\begin{equation}\label{eqn:lig}
    h(x) = \sum_{i=0}^{n-1}a_ih_i(x) = a_0h_0(x) + a_1h_1(x) + \dots + a_{n-1}h_{n-1}(x)
\end{equation}

 where $x$ is the Cartesian coordinates with the midpoint of the two circles as the origin of the coordinates, and $h_0(x), h_1(x), \dots, h_{n-1}(x)$ are continuous and pre-selected functions. To ensure the smoothness of the ligament, we require the the zeroth, first, and second derivatives of the $h(x)$ to be continuous at the two ends M and N.   The coordinates of M are represented as $(x_M, y_M)$ and N as $(x_N, y_N)$. Consequently, we have six known constraints on $h(x)$:
 
\begin{equation}\label{eqn:boundary}
    \begin{cases}
    h(x_M) = y_M  \\
    h^{'}(x_M) = y_M^{'}\\
    h^{''}(x_M) = y_M^{''} \\ 
    h(x_N) = y_N \\
    h^{'}(x_N) = y_N^{'}\\
    h^{''}(x_N) = y_N^{''} \\ 
    \end{cases}
\end{equation}

where $y_M, y_M^{'}$, and $y_M^{''}$ denote the zeroth, first, and second derivatives at point M, and $y_N, y_N^{'}$, and $y_N^{''}$ denote the zeroth, first, and second derivatives at point N. Given that the midpoint of the two circles is the origin, the center of the left circle is at $(-L/2,0)$, and the center of the right circle is at $(L/2,0)$. Thus, we have:

\begin{equation}
    \label{eqn:coords}
    \begin{cases}
    x_M = -L/2+ (R+2R_s)\cdot \cos(\theta_0) \\
    y_M = (R+2R_s)\cdot \sin(\theta_0) \\
    y_M^{'} = -1/\tan(\theta_0)\\
    y_M^{''} = -R^2/y_M^3  \\
    x_N = L/2 + (R+2R_s)\cdot \cos(\theta_{00}) \\
    y_N = (R+2R_s)\cdot \sin(\theta_{00}) \\
    y_N^{'} = -1/\tan(\theta_{00})\\
    y_N^{''} = -R^2/y_N^3 \\ 
    \end{cases}
\end{equation}

Consider that the number of coefficients to be determined for $h(x)$ is $n$ and we only have $6$ boundary conditions in Eq. \ref{eqn:boundary}. By solving Eq. \ref{eqn:lig} with Eq. \ref{eqn:boundary}, $h(x)$ has a unique solution when $n$ is equal to $6$ and multiple solutions when $n$ exceeds $6$. In this work, we set n to 7 to add one more degree of freedom to the shape of the ligament by fixing a randomly selected point $(x^*,y^*)$ on the ligament, denoted as a red star point in Fig. \ref{fig:ds}(a). Then we have one more constraint: 
\begin{equation}\label{eqn:random}
    h(x^*) = y^*
\end{equation}

The coefficients $a_0, a_1, \dots, a_6$ of $h(x)$ can be identified by Eq. \ref{eqn:boundary} and Eq. \ref{eqn:random}.

In summary, we have $8$ design parameters $X = [R, \theta_0,  \theta_1,  \theta_{00}, \theta_{11},  R_s, x^*, y^*]$. Notably, $R_s$ can take the value of zero, and the subfunction components $h_0(x),h_1(x), \dots, h_6(x)$ of the ligament can be either polynomial or trigonometric.  To search for optimal designs for the objectives defined in Section \ref{sec:obj}, we explore four different design spaces, as illustrated in Fig. \ref{fig:ds}(b). The range of the design parameters [$R, \theta_0,  \theta_1,  \theta_{00}, \theta_{11}$ ] of each design spaces are presented in Table \ref{tab:ds}. The details of the value of $R_s$ and the sampling function $h(x)$ for the ligament in each design space are summarized below: 

\begin{enumerate}
    \item Design Space 1
    \begin{itemize}
        \item $R_s = 0$ for both left and right side
        \item $h(x) = \sum_{i=0}^{n-1}a_ix^i$
    \end{itemize}
    \item Design Space 2
    \begin{itemize}
        \item $0.1\leq R_s \leq R/10$ for both left and right side
        \item $h(x) = \sum_{i=0}^{n-1}a_ix^i$
    \end{itemize}
    \item Design Space 3
    \begin{itemize}
        \item $R_s = 0$ for both left and right side
        \item $h(x) = a_0 + a_1x + a_2sin(x) + a_3sin(x) + a_4x^4 + a_5x^5 + a_6x^6$. 
    \end{itemize}
    \item Design Space 4
    \begin{itemize}
        \item $R_s = 0$ for right side and $0.1\leq R_s \leq R/10$ for left side
        \item $h(x) = \sum_{i=0}^{n-1}a_ix^i$
    \end{itemize}
\end{enumerate}

As discussed in Section \ref{sec:contact}, diverse contact mechanisms can induce varying non-reciprocity and asymmetry properties. The choice of the four design spaces shown in Fig. \ref{fig:ds}(b) aims to encompass a broad range of contact mechanisms. In design space 1, the ligament and the right circle are prone to establish contact when it is subjected to extension. In design space 2, the ligament and the right circle tend to make contact under compression. Design space 3 introduces alterations to the subfunctions of the ligament shape compared to design space 1, allowing us to investigate the impact of the ligament shape on the properties of the chiral metamaterial. In design space 4, the left circle and the ligament tend to contact under compression, while the right circle and the ligament tend to contact under extension loads.

\begin{table}[]  
\centering
\caption{Design parameter ranges within each design space.}
\begin{tabular}{@{}cccccc@{}}
\toprule
 & R & $\theta_0$ & $\theta_1$ & $\theta_{00}$ & $\theta_{11}$ \\ \midrule
Design Space 1 & {[}3,7{]} & {[}-90, -20{]} & {[}-90, -20{]} & {[}90, 160{]} & {[}-90, -20{]} \\ \midrule
Design Space 2 & {[}3,7{]} & {[}-90, 0{]} & {[}20, 90{]} & \multicolumn{1}{l}{{[}90,180{]}} & \multicolumn{1}{l}{{[}20,90{]}} \\ \midrule
Design Space 3 & {[}3,7{]} & {[}-90, -20{]} & {[}-90, -20{]} & {[}90, 160{]} & {[}-90, -20{]} \\ \midrule
Design Space 4 & {[}3,7{]} & {[}-90, 0{]} & {[}20, 90{]} & \multicolumn{1}{l}{{[}90,180{]}} & \multicolumn{1}{l}{{[}-90,-20{]}} \\ \bottomrule
\label{tab:ds}
\end{tabular}
\end{table}

\section{Machine Learning Methods}
\label{sec:method}
Machine learning (ML) methods have been extensively applied to complex engineering challenges, demonstrating their efficiency in uncovering intricate relationships within data ~\citep{lee2023data,ha2023rapid,buehler2023melm,yuan2022towards,arzani2023interpreting,nguyen2024segmenting}. Through sufficient training on diverse datasets, an effectively trained ML model exhibits rapid and accurate predictions on unseen data, drastically reducing the time required to characterize material properties by orders of magnitude compared to conventional experimental or simulation methods. This capacity enables it to efficiently explore a large pool of candidates in the search for optimal designs. In the context of chiral metamaterial design, the design space is practically infinite because the design parameters governing the initial contact angle and the shape of the ligament shape are continuous. Consequently, we employed ML methods to guide our exploration of nonreciprocal and asymmetric stiffness chiral metamaterials. The following parts of this section elaborate on our approach to data collection, the details of the ML model, and the data-driven optimization methods utilized to discover optimal designs. 

\subsection{Data Collection}
\label{sec:data}

\subsubsection{Data Representation}

\begin{figure}[h!]
    \centering
    \includegraphics[width=.9\textwidth]{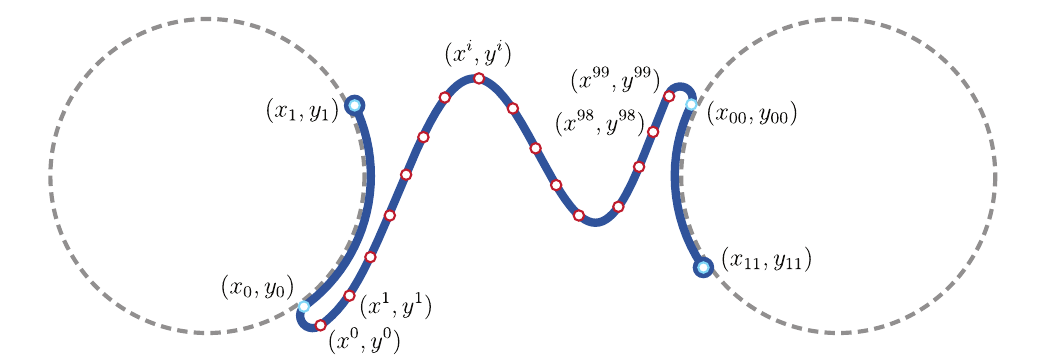}
    \caption{Illustration of the feature representation of a chiral metamaterial}
    \label{fig:feature}
\end{figure}

In Section \ref{sec:ds}, we outlined each design space using eight design parameters, denoted as $X = [R, \theta_0,  \theta_1,  \theta_{00}, \theta_{11},  R_s, x^*, y^*]$. However, due to the diversity in the subfunctions of $h(x)$ that define the ligament shape, these parameters cannot fully characterize the geometry of the chiral structure across all design spaces. Additionally, given the efficacy of machine learning methods in pattern recognition~\citep{bishop2006pattern}, we aim to represent the chiral structure focusing on its shape and geometric features. As Fig. \ref{fig:feature} shows, we note the points where the ligament and the rigid circles initiate connection as $(x_0,y_0)$ and $(x_{00},y_{00})$, and coordinates of the two ends of the ligament as $(x_1,y_1)$ and $(x_{11},y_{11})$. Furthermore, we uniformly sample $100$ points along the $x$ direction from the elastic ligament. Therefore, we employed the coordinates of $104$ points to uniquely describe the geometry of each chiral structure. The $208$ features of a single chiral structure are summarized as feature vector $X$ as follows:
 
\begin{equation}
    X=[x^0, \, x^1,\dots, \, x^{99}, \, y^0, \, y^1,\dots,\, x^{99},\, , \, x_1, \, y_1, \, x_0, \, y_0, \, x_{00}, \, y_{00}, \, x_{11}, \, y_{11}]
\end{equation}

\begin{figure}[h!]
    \centering
    \includegraphics[width=.9\textwidth]{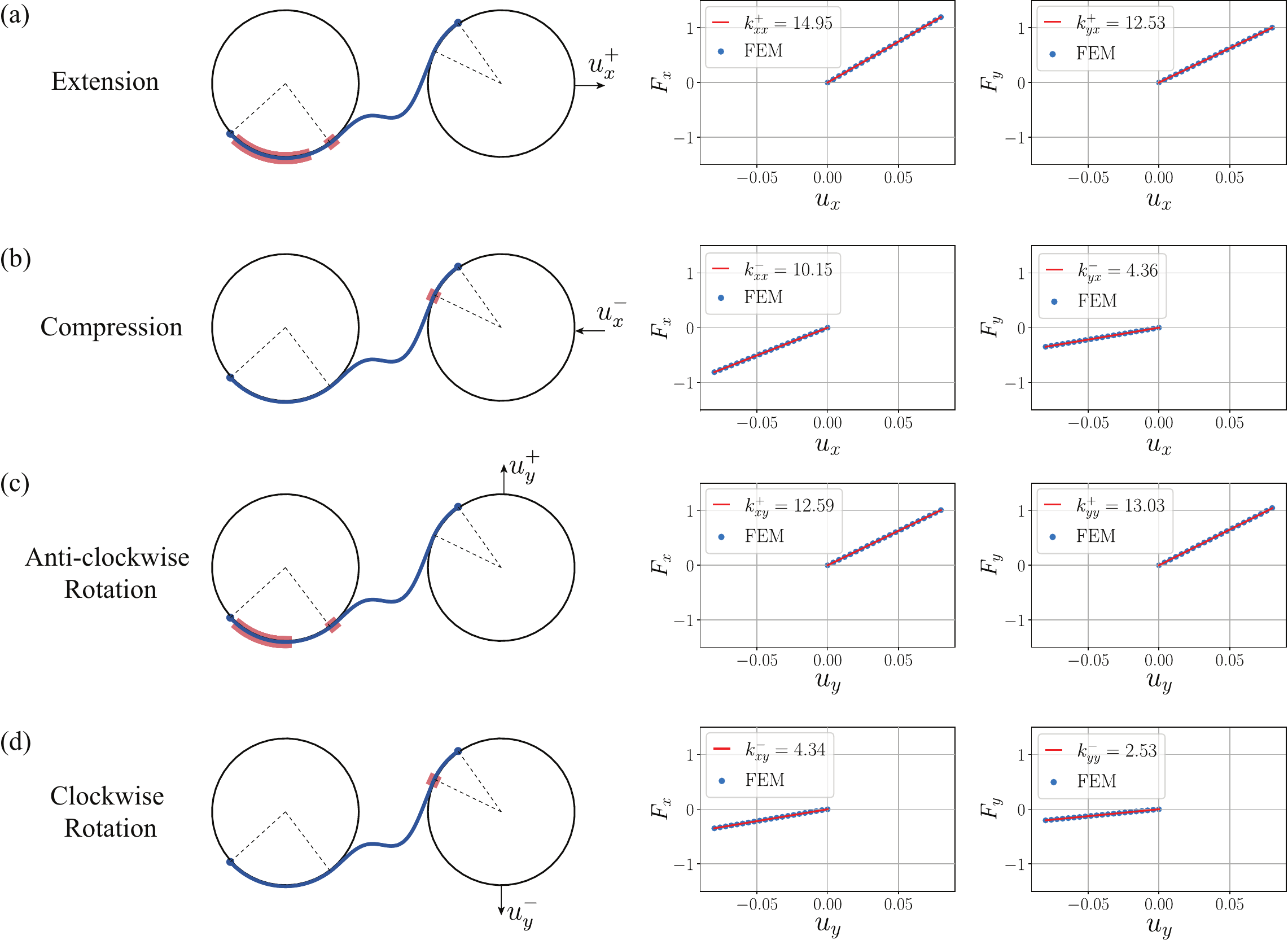}
    \caption{Illustration of a representative chiral structure subjected to loads from four different directions and the corresponding force-displacement response obtained from finite element (FEM) simulation. The contact area is highlighted in pink. Stiffness is determined as the coefficient of linear fitting applied to the FEM data using the least squares method.}
    \label{fig:obj}
\end{figure}

The second step is to label each chiral structure with its mechanical properties by running a finite element simulation. As the goal of this study is to maximize the nonreciprocity and asymmetry defined in Section \ref{sec:obj}, the label of each material will be the objectives $f_{1:8}$ and $g_{1:8}$ defined in Section \ref{sec:obj_nonreci} and Section \ref{sec:obj_asym}. To get this, it is necessary to first calculate the stiffness values for each chiral structure. Fig. \ref{fig:obj} provides an example of how we obtained the stiffness values of each structure. Each row shows the contact region of the structure under the loads from four different directions and the corresponding force-displacement response. The stiffness, i.e. the slope of the response curve, is obtained by fitting the finite element simulation results using least squares. 

In Fig. \ref{fig:obj}(a)(b), due to the different contact regimes in compression and extension, we obtain $k_{xx}^+ = 14.95$ and $k_{yx}^+ = 12.53$ for extension, and $k_{xx}^- = 10.15$ and $k_{yx}^- = 4.36$ for compression. In Fig. \ref{fig:obj}(c) where the structure is subject to an anti-clockwise rotation load, we obtain  $k_{yy}^+ = 13.03$ and $k_{xy}^+ = 12.59$. Notably, the contact regime under the anti-clockwise rotation load in Fig. \ref{fig:obj}(c) closely resembles that of extension in Fig. \ref{fig:obj}(a), which results in the values of $k_{xy}^+ = 12.59$ and $k_{yx}^+ = 12.53$ being nearly equal.  In Fig. \ref{fig:obj}(d) where the structure is subject to a clockwise rotation load, we obtain $k_{yy}^- = 2.53$ and $k_{xy}^- = 4.34$. Similarly, the contact mode of clockwise rotation in Fig. \ref{fig:obj}(d) closely resembles the compression mode in Fig. \ref{fig:obj}(b), and $k_{xy}^- = 4.34$ is nearly equal to $k_{yx}^- = 4.36$. These observations align with the discussion of the contact mechanism in Section \ref{sec:contact}, indicating that the contact modes play a pivotal role in governing the non-reciprocity and asymmetry properties of the chiral structures. Subsequently, we can identify the non-reciprocity and asymmetry objectives $f_{1:8}$ and $g_{1:8}$ defined in Section \ref{sec:obj_nonreci} and Section \ref{sec:obj_asym} for this design in Fig. \ref{fig:obj}. The value are listed below:

$f_1 = |\frac{k_{xx}^-}{k_{xx}^+}| = |10.15/14.95| = 0.67$, \, 
$f_2 = |\frac{k_{xx}^+}{k_{xx}^-}| = |14.95/10.15| = 1.47$, \,
$f_3 = |\frac{k_{xy}^-}{k_{xy}^+}|= |4.34/12.59| = 0.34$, 

$f_4 = |\frac{k_{xy}^+}{k_{xy}^-}| = |12.59/4.34| = 2.90$, \,
$f_5 = |\frac{k_{yx}^-}{k_{yx}^+}| = |4.36/12.53| = 0.34$, \,
$f_6 = |\frac{k_{yx}^+}{k_{yx}^-}| = |12.53/4.36| = 2.87$, 

$f_7 = |\frac{k_{yy}^-}{k_{yy}^+}| = |2.53/13.03| = 0.19$, \,
$f_8 = |\frac{k_{yy}^+}{k_{yy}^-}| = |13.03/2.53| = 5.15$,\,
$g_1 = |\frac{k_{xy}^-}{k_{yx}^-}| = |4.34/4.36| = 0.99$, 

$g_2 = |\frac{k_{xy}^-}{k_{yx}^+}| = |4.34/12.53| = 0.34$, \,
$g_3 = |\frac{k_{xy}^+}{k_{yx}^-}| = |12.59/4.36| = 2.88$, \,
$g_4 = |\frac{k_{xy}^+}{k_{yx}^+}| = |12.59/12.53| = 1.00$, 

$g_5 = |\frac{k_{yx}^-}{k_{xy}^-}| = |4.36/4.34| = 1.00$, \,
$g_6 = |\frac{k_{yx}^+}{k_{xy}^-}| = |12.53/4.34| = 2.88$, \,
$g_7 = |\frac{k_{yx}^-}{k_{xy}^+}| = |4.36/12.59| = 0.34$,

$g_8 = |\frac{k_{yx}^+}{k_{xy}^+}| = |12.53/12.59| = 0.99$.

\subsubsection{Data Augmentation}
\begin{figure}[h!]
    \centering
    \includegraphics[width=.9\textwidth]{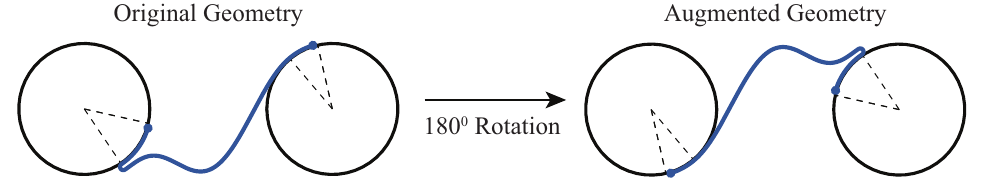}
    \caption{Illustration of the rotation transformation for data augmentation. The augmented geometry is obtained by rotating the original geometry by 180 degrees. The original and the augmented structure have the same properties in terms of their stiffness values in all directions.}
    \label{fig:aug}
\end{figure}

Data augmentation is a technique used to increase the diversity and the size of the training dataset.  New training samples are generated by applying transformations or modifications to existing data, such as flipping, rotating, and scaling \citep{shorten2019survey, rebuffi2021data}. Given the computational expense of acquiring mechanical properties via finite element simulation, data augmentation becomes essential for chiral metamaterial design. Across all design spaces, the stiffness properties of a chiral material remain invariant under a 180-degree rotation, as Fig. \ref{fig:aug} shows. Leveraging this property, we apply the rotation transformation to each of the chiral metamaterial data points and double the size of the dataset.

\subsection{Bayesian Optimization}
\label{sec:bo}
During the search for optimal chiral structures, obtaining properties such as non-reciprocity and asymmetry by FEM is a challenging and computationally expensive task. Moreover, the physics of the relationship between the design and the properties of the chiral structure is nonlinear, which makes it challenging to obtain analytic solutions for the optimization objective. To address these challenges, we propose leveraging a data-driven approach - Bayesian Optimization~\citep{polyzos2023bayesian}. This method is a model-based optimization technique that excels in efficiently determining optimal designs for objective functions that are black-box and costly to evaluate. In the framework of material design, the goal of Bayesian Optimization is to address the optimization problem presented below:

\begin{equation}\label{eq:bo}
    x^* = arg \max_{x \in \mathcal{X}} f(x)
\end{equation}

where $\mathcal{X}$ compromise all feasible designs, and $f(x)$ represents the objective function to be maximized. For each design point $x$ that belongs to $\mathcal{X}$, the goal is to identify the optimal point $x^*$ at which the objective $f(x)$ attains its maximum value. To achieve this, we must select a probabilistic surrogate model $f(x)$ and an acquisition function $\alpha(x)$.  The surrogate model $f(x)$ evaluates the objective value of each design point, and the acquisition function $\alpha(x)$ determines which point to query next. The query process involves obtaining the ground truth property for the selected point through experiments or simulation tools. 

To conduct Bayesian Optimization, we start with an observed data collection $D_t =\{(x_i,y_i)\}_{i=1}^t$, where $y_i$ denotes the measured objective value corresponding to point $x_i$. The first step is to fit the surrogate model with the dataset $D_t$. The second step is selecting the next point $x_{t+1}$ that maximizes the acquisition function, querying the value of the $y_{t+1}$, and adding $(x_{t+1}, y_{t+1})$ to the observed dataset. The final solution to Eq. \ref{eq:bo} is obtained by repeating these two steps until reaching a maximum iteration limit $N$. Below is a pseudo-code outlining the process of Optimization Routine using Bayesian Optimization:

\begin{algorithm}
\caption{Optimization Routine}
\label{alg:bo}
\begin{algorithmic}[1]
\Require initial observations $D_0$, maximum iterations $N$, surrogate model $f(x)$, acquisition function $\alpha (x)$
\For {$t = 0, 1 ,2, \dots, N-1$}
\State fit the model $f(x)$ with $D_t$
\State calculate the acquisition function $\alpha (x)$ with $f(x)$
\State select the next point $x_{t+1}  \gets arg \max_x \alpha (x)$
\State perform experiment or simulation to evaluate the objective values $y_{t+1}$ at point $x_{t+1}$ 
\State $D_{t+1} \gets \{D_t,(x_{t+1},y_{t+1})\}$ 
\EndFor 
\State \Return best $y$ and the corresponding $x$ in $D_N$
\end{algorithmic}
\end{algorithm}

When employing Bayesian Optimization for chiral metamaterial design, we begin by conducting simulations for $5$ randomly selected designs in each of the four design spaces. Thus we establish an initial dataset with a total of $20$ points. We then generate a pool of $400,000$ design candidates, with $100,000$ in each design space, without conducting FEM simulation. Subsequently, we perform the Bayesian Optimization for $10$ iterations. In each iteration, the top $10$ points in the design pool with the highest values determined by the acquisition function are selected for running FEM simulations using Abaqus. This approach of acquiring ten points rather than one in each iteration is adopted due to our simulation setup, where multiple simulation jobs are submitted as a batch in the computing center. Therefore, querying $10$ points per iteration takes the same amount of time as querying a single point. Moreover, expanding the dataset faster enables us to get a more accurate surrogate model, expediting the discovery of optimal designs.

\subsubsection{Surrogate Model}
The Gaussian Process (GP) model~\citep{schulz2018tutorial,deringer2021gaussian,wang2023intuitive} is commonly favored as a surrogate model choice in Bayesian Optimization due to its capability to estimate prediction uncertainty directly. However, its efficacy diminishes notably in high-dimensional problems~\citep{ shahriari2015taking, binois2022survey}. Additionally, the intuition behind GP begins with kernel-encoded prior assumptions about the dataset distribution, then obtains a posterior distribution function given the observed data \citep{marrel2024probabilistic, schulz2018tutorial}. However, in the case of intricate chiral material systems, where prior dataset knowledge is absent and the feature dimension is high, extra challenges arise when selecting appropriate kernels for GP regression models. In contrast, deep learning models like Multilayer Perceptron (MLP) are efficient at capturing black-box, unknown nonlinear relationships within high-dimensional data. Nonetheless, MLP models do not inherently provide uncertainty estimates for predictions. To address this, we employ ensemble learning methods~\citep{ganaie2022ensemble,mienye2022survey,mohammadzadeh2023investigating} to train a single MLP model multiple times with different seeds on observed data and utilize the variance across model predictions as an indication of uncertainty.

The MLP model is designed with four layers comprising 1024, 1024, 64, and 1 neurons, respectively. Following each hidden layer, Rectified Linear Unit (ReLU) activation functions were applied to introduce non-linearity into the model. The training was performed using a batch size of 16 over 250 epochs, with a learning rate set to 0.001. We train the MLP model $k = 15$ times, each utilizing a different seed for training-validation dataset splitting and weight initialization. The prediction of objective values and the uncertain estimation are calculated below: 

\begin{equation}
    \begin{cases}
        \mu (x) = \frac{1}{k} \sum_1^k \hat y_i \\
        \sigma (x) = \sqrt{\frac{1}{k} \sum_1^k (\hat y_i - \mu (x))^2}
    \end{cases}
    \label{eq:prediction}
\end{equation}

where $\mu (x)$ is the averaged prediction across all MLP models,  $\sigma (x)$ is the standard deviation of the predictions, and $\hat y_i$ is the predicted value from the $i$th model. 

\subsubsection{Acquisition Function}
In the framework of Bayesian Optimization, various acquisition functions are available for selection, such as the probability of improvement (PI)~\citep{kushner1964new, ruan2020variable}, entropy search (ES)~\citep{hennig2012entropy,wang2017max}, expected improvement (EI)~\citep{zhan2020expected,qin2017improving}, and upper confidence bound (UCB)~\citep{carpentier2011upper,simchi2023multi}. In this paper, we focus on introducing two specific methods we employed in our study: EI and UCB.

\textbf{Expected Improvement}

The Expected Improvement (EI) method is a versatile choice for many optimization scenarios. It is efficiently designed to trade-off between the global search (exploration) and local minimization (exploitation) ~\citep{brochu2010tutorial}. The acquisition function of EI is:

\begin{equation}\label{eq:EI0}
    \alpha^{\textrm{EI}} (x) = \mathrm{E} [\max (0, f(x) - f_{\max})]
\end{equation}

where $f_{\max}$ is the best performance observed so far. The acquisition function $\alpha^{\textrm{EI}} (x)$ calculates the expectation of the improvement for each unseen point. In essence, the equation evaluates each point by how much performance it can enhance compared to the current best point.  If the potential improvement surpasses the best point, the expected improvement is the amount of improvement. If the potential improvement cannot surpass the best point, the expected improvement is zero. Assuming the target $y$ follows a Gaussian distribution $\mathcal{N}(\mu(x),\sigma(x) )$, the formulation of EI can be written explicitly as ~\citep{jones1998efficient, brochu2010tutorial}

\begin{equation}\label{eq:EI}
    \alpha^{\textrm{EI}} (x) = (\mu(x)-f_{\max}-\xi) \Phi(\lambda)+\sigma(x) \phi(\lambda)
\end{equation}

where  $\lambda= (\mu(x)-f_{\max}-\xi)  / \sigma(x)$. $\Phi(\lambda)$ and $\phi(\lambda)$ are the cumulative distribution function (CDF) and probability density function (PDF) of the standard normal distribution respectively. The first term in Eq. \ref{eq:EI} emphasizes exploitation by favoring points with higher expected objective values, while the second term promotes exploration by favoring points with greater uncertainty. Thus, Eq. \ref{eq:EI} strikes a balance between exploitation and exploration controlled by the hyperparameter $\xi$, with higher $\xi$ leading to more exploration. In our study, we set $\xi$ to $0.002$ to achieve a balanced exploration-exploitation strategy.

\textbf{Upper Confidence Bound}

The Upper Confidence Bound (UCB) calculated the upper bound for the prediction by adding the uncertainty to the estimation, offering a straightforward approach to balancing exploration and exploitation. The acquisition function for UCB is defined as ~\citep{shahriari2015taking, cox1992statistical}:

\begin{equation}\label{eq:ucb}
    \alpha^{\textrm{UCB}} (x) = \mu(x) + \beta \sigma(x)
\end{equation}

Similar to EI, the first component of the equation favors exploitation, and the second component for exploration. The hyperparameter $\beta$ serves to balance the weights between exploitation and exploration, with higher values leading to more exploration. 

In our problem, we apply EI to single objective optimization considering that the algorithm has already proven efficient in material discovery in many other studies~\citep{gongora2020bayesian,zhang2020bayesian,kotthoff2021bayesian}. Although the EI acquisition function form of Eq. \ref{eq:EI} is typically associated with the Gaussian Process Model, we adopted the acquisition function Eq. \ref{eq:EI} to our problem with the assumption that the observed data are sampled from a Gaussian Distribution with mean and variance derived from the predictions of ensembled MLPs, i.e. $y \sim \mathcal{N}(\mu,\sigma)$, as formalized in Eq. \ref{eq:prediction}. This assumption is grounded in the evidence that MLPs (or Neural Networks) are well-calibrated models~\citep{niculescu2005predicting}, thus allowing us to treat the mean and variance of prediction as the posterior distribution of objectives. While the detailed exploration of uncertainty estimation of deep ensembled regression models is not the primary emphasis of this paper, readers interested in a more comprehensive and rigorous discussion on this topic can refer to the literature ~\citep{lakshminarayanan2017simple, nix1994estimating, abdar2021review}.

However, in the scenario of multi-objective optimization, fewer valid candidates meet the criterion due to the increased number of objectives to optimize and all objectives have to be larger than $1$ (refer to Section \ref{sec:obj_multi}). The EI method is unable to filter out invalid designs in this context. Hence, we employ the Upper Confidence Bound (UCB) acquisition method to guide the multi-objective optimization. The acquisition value calculated by UCB acts as an upper bound for the objective value, enabling us to disregard candidates whose objective values fall below $1$. The hyperparameter $\beta$ is set to $3$ to increase the upper confidence bound for potential optimal designs, thereby allowing more valid candidates to be considered.

\subsubsection{Pareto Front}
In Section \ref{sec:obj_multi}, we outlined our approach to addressing the multi-objective optimization challenge in chiral metamaterial design by searching for the Pareto front across all designs. In this Section, we elaborate on the methodology of discovering the Pareto front and its utilization within Bayesian Optimization. During each iteration of multi-objective Bayesian optimization, we determine the acquisition value of each unseen data point using the Upper Confidence Bound (UCB) method. Provided that we aim to optimize a total of $m$ objectives simultaneously, the acquisition value of points $i$, calculated by Eq. \ref{eq:ucb}, is denoted as $\alpha^i = [\alpha_1^i, \alpha_2^i, \dots, \alpha_m^i]$, where $\alpha_j^i$ represents the acquisition value for the unseen data point $i$ in terms of maximizing objective $j$. Therefore, if we have $n$ unseen points in the pool of design candidates, we can obtain a matrix $A_{n\times m}$ with row $i$ representing $\alpha^i$, where $i\in \{1,2, \cdots, n\}$. The next points to be acquired will be the design points whose performances form the Pareto front of $A_{n\times m}$. The Pareto front is identified through an efficient algorithm available in the Python package \textbf{artemis-ml} \citep{quva_lab_artemis}. The algorithm iteratively eliminates the points that are dominated by at least one other point until only non-dominated points remain, thus yielding the set of Pareto front. The algorithm can be summarized as follows:

\begin{algorithm}
\caption{Find Pareto Front}
\label{alg:pareto}
\begin{algorithmic}[1]
\Require Acquisition value $A_{n\times m}$ for $m$ objectives of unseen points $1,2,\cdots\,n$
\State choose the first point $i_1 = 1$
\While {$i_1<=n$}
\For {$i_2 = 1 ,2, \dots, n$}
\State remove point $i_2$ if $c_j^{i_2}<c_j^{i_1}$ for all $j \in \{1,2,\cdots,m\}$ \Comment{$i_2$ is removed because it is dominated by $i_1$} 
\EndFor
\State update $n$ as the size of the rest points
\State re-index the rest points as $1,2,\cdots,n$  without changing the order
\State $i_1 \gets$ the index of the next point on the order 
\EndWhile
\State \Return the rest points (i.e. nondominated points)
\end{algorithmic}
\end{algorithm}

\section{Results and Discussion}
\label{sec:res}
In this Section, we present the optimal designs discovered using the Bayesian Optimization detailed in Section \ref{sec:bo}. In Section \ref{sec:res_nonreci}, we present the results and analysis for the optimal designs that exhibit extreme non-reciprocity. In Section \ref{sec:res_asym}, we present the results and analysis for the optimal designs that exhibit extreme elastic asymmetry. Finally, in Section \ref{sec:res_multi}, we present optimal designs that were discovered by maximizing both elastic asymmetry and non-reciprocity. 

\subsection{Non-Reciprocity Optimization}
\label{sec:res_nonreci}

\begin{figure}[h!]
    \centering
    \includegraphics[width=.9\textwidth]{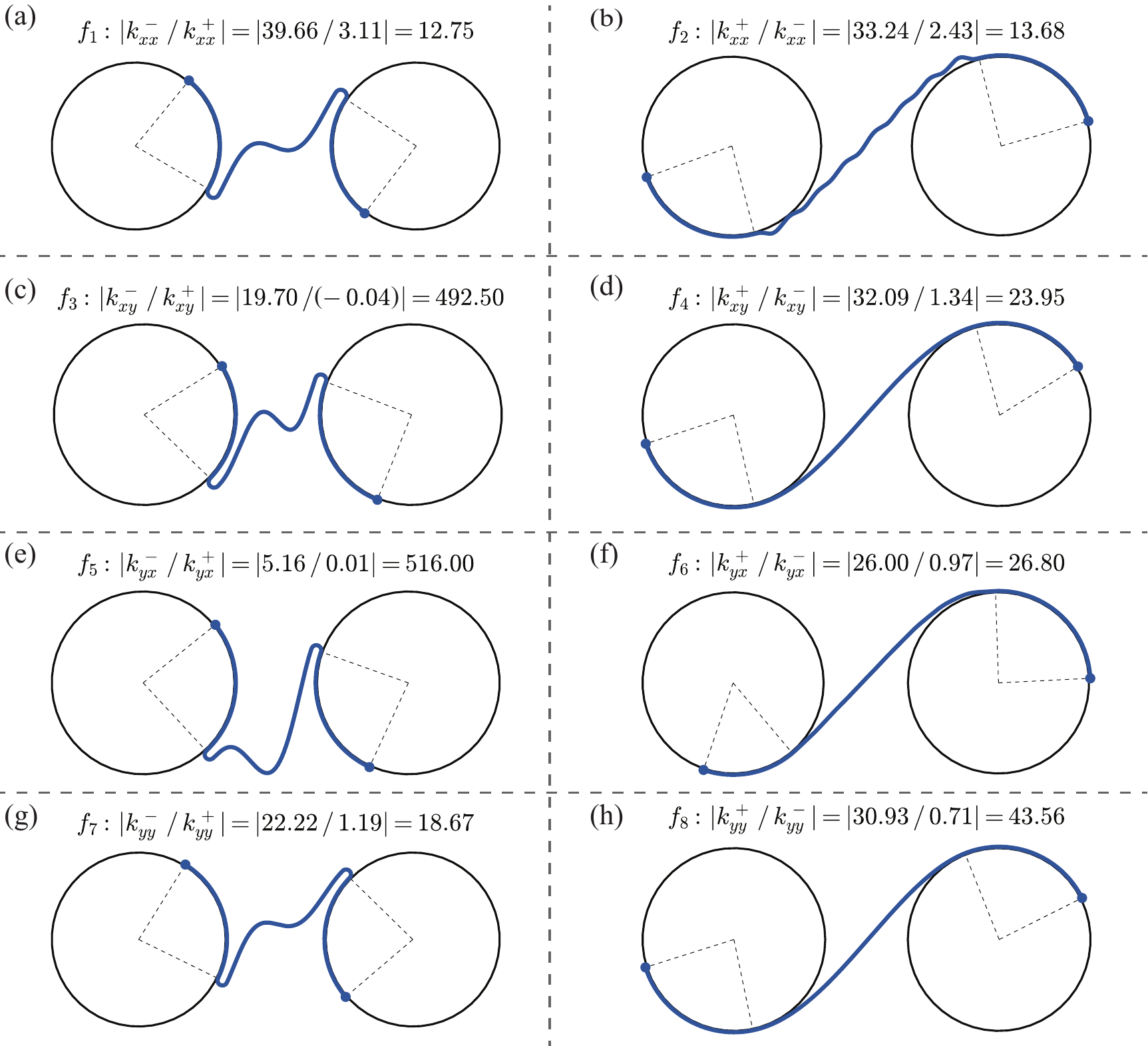}
    \caption{Optimal designs for maximizing eight non-reciprocity objectives after $10$ iterations of Bayesian Optimization. Each figure shows the optimal design for the (a) objective $f_1$, (b) objective $f_2$, (c) objective $f_3$, (d) objective $f_4$, (e) objective $f_5$, (f) objective $f_6$, (g) objective $f_7$ and (h) objective $f_8$. The title above each structure indicates the corresponding stiffness values.}
    \label{fig:res_nonreci}
\end{figure}

For the eight non-reciprocity objectives defined in Section \ref{sec:obj_nonreci}, we search for the optimal design that maximizes each objective by conducting the Bayesian Optimization methods detailed in Section \ref{sec:bo}. Each objective is maximized individually through separate Bayesian Optimization runs. The designs achieving the highest objective value after optimization are shown in Fig. \ref{fig:res_nonreci}. We present a summary of the optimal performance for each objective below

$f_1 = |\frac{k_{xx}^-}{k_{xx}^+}| = |39.66/3.11| = 12.75$, \,
$f_2 = |\frac{k_{xx}^+}{k_{xx}^-}| = |33.24/2.43| = 13.68$, \,

$f_3 = |\frac{k_{xy}^-}{k_{xy}^+}| = |19.70/(-0.04)| = 492.50$, \,
$f_4 = |\frac{k_{xy}^+}{k_{xy}^-}| = |32.09/1.34| = 23.95$, \,

$f_5 = |\frac{k_{yx}^-}{k_{yx}^+}| = |5.16/0.01| = 516.00$, \,
$f_6 = |\frac{k_{yx}^+}{k_{yx}^-}| = |26.00/0.97| = 26.80$, \,

$f_7 = |\frac{k_{yy}^-}{k_{yy}^+}| = |22.22/1.19| = 18.67$, \,
$f_8 = |\frac{k_{yy}^+}{k_{yy}^-}| = |30.93/0.71| = 43.56$, \,

\begin{figure}[h!]
    \centering
    \includegraphics[width=.9\textwidth]{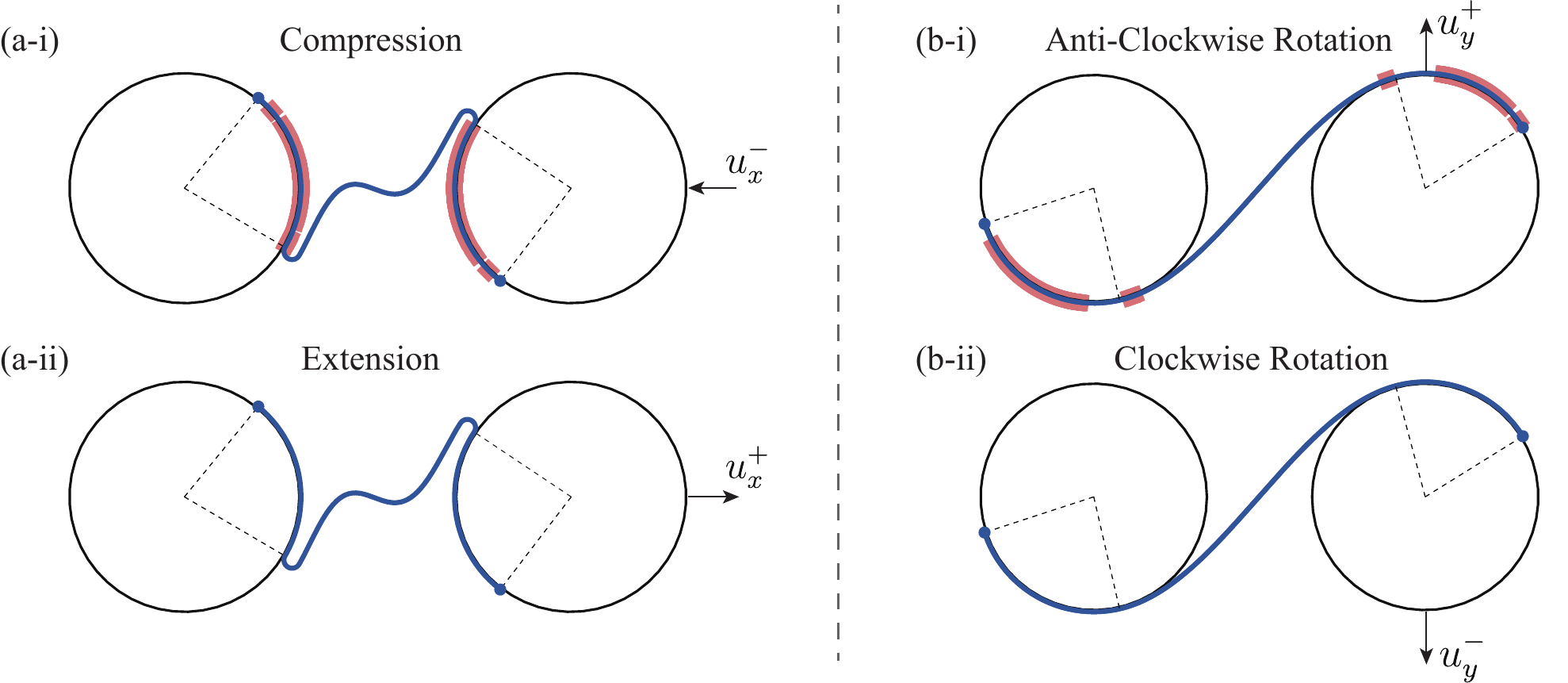}
    \caption{Two examples of contact modes under loads from different directions. The highlighted pink color indicates the contact areas after the deformation. (a) The contact modes for the optimal design of objective $f_1= |k_{xx}^-/k_{xx}^+|$.  The stiffness value $k_{xx}^-$ is obtained during the (a-i) compression, and $k_{xx}^+$ is obtained during the (a-ii) extension. (b) The contact modes optimal design of objective $f_4 = |k_{yy}^-/k_{yy}^+|$. The stiffness value $k_{yy}^+$ is obtained during the (b-i) anti-clockwise rotation, and $k_{yy}^-$ is obtained during the (b-ii) clockwise rotation.}
    \label{fig:res_nonreci_contact}
\end{figure}

It is clear that the optimal designs for objectives $f_1, f_3, f_5, f_7$ in Fig. \ref{fig:res_nonreci}(a)(c)(e)(g) share some similarities, and are all discovered from design space 2. Referring to the stiffness definition illustrated in Fig. \ref{fig:stiff}, these objectives necessitate either that the stiffness values for compression ($-x$) are greater than those for extension ($+x$), or the stiffness values for clockwise rotation ($-y$) are greater than those for anti-clockwise rotation ($+y$). By looking into the deformation process of these structures under various loads, we observe that during compression ($-x$) and clockwise rotation ($-y$), contact is established between the ligament and the rigid circle. Conversely, during extension and anti-clockwise rotation, the ligament and the rigid circle detach, resulting in no contact forces during deformation. This mechanism can be understood through intuitively imaging the deformation process of these geometries. Here, we select the optimal structure for objective $f_1 = |k_{xx}^-/k_{xx}^+|$ as an example and show its contact behavior during deformation in Fig. \ref{fig:res_nonreci_contact}(a). Notably, there exists a substantial contact area when the structure is subject to compression load (see Fig. \ref{fig:res_nonreci_contact}(a-i)) and no contact area for the extension load (see Fig. \ref{fig:res_nonreci_contact}(a-ii)). The optimal designs for objectives $f_3, f_5$, and $f_7$ exhibit similar contact modes, which are visualized in \ref{apx:contact}, Fig. \ref{fig:apx_nonreci}.

Essentially, higher stiffness values are typically observed during contact, as illustrated in Fig. \ref{fig:res_nonreci_contact} and Fig. \ref{fig:apx_nonreci}, while lower stiffness values are observed when there is no contact. Additionally, based on the design parameters outlined in Table \ref{tab:ds}, the initial contact angles between the ligament and the circle, denoted as $\theta_{1}$ and $\theta_{11}$, range from $20$ to $90$ degrees. Notably, the initial contact angles of the optimal designs depicted in Fig. \ref{fig:res_nonreci} (a)(c)(e)(g) are all closer to the upper bound, i.e., $90$ degrees, which enables the establishment of a larger contact area under loads. Therefore, maximizing the difference between higher and lower stiffness promotes a larger initial contact area between the ligament and the rigid circle. Upon closer inspection of the optimal designs, it is evident that the area where the ligament and the circle connect is large and towards the center of the structure provides stronger resistance during compression loads. Thus, we can see that the larger initial connecting area between the ligament and the circle tends to enhance the stiffness, as it will increase the contact area under loads from certain directions. Moreover, the shape of the ligament for the designs depicted in Fig. \ref{fig:res_nonreci}(a)(c)(e)(g) tends to be curved in the middle, which provides additional resistance against the compression load.

The optimal designs for objectives $f_2, f_4, f_6, f_8$ in Fig. \ref{fig:res_nonreci}(b)(d)(f)(h) represent the reversed versions of those in Fig. \ref{fig:res_nonreci}(a)(c)(e)(g). Specifically, for the four objectives $f_2, f_4, f_6$, and $f_8$, the goal is to design material such that the stiffness values for extension ($+x$) are greater than those for compression ($-x$), and the stiffness values for anti-clockwise rotation ($+y$) are greater than those for clockwise rotation ($-y$). Accordingly, the optimal geometries for these objectives in Fig. \ref{fig:res_nonreci}(b)(d)(f)(h) are also alike but in contrast with those of the optimal designs in Fig. \ref{fig:res_nonreci}(a)(c)(e)(g). Concretely, in this set of optimal designs, the contact areas are situated away from the central region, and the ligaments are straight rather than curved. In addition, contact occurs during extension and anti-clockwise rotation, while no contact is observed during compression and clockwise rotation, which contrasts with the structures in Fig. \ref{fig:res_nonreci}(a)(c)(e)(g). We present the contact modes of the optimal structure for objective $f_4 = |k_{yy}^-/k_{yy}^+|$ in Fig. \ref{fig:res_nonreci_contact}(b). Here, we observe a substantial contact area during anti-clockwise rotation (see Fig. \ref{fig:res_nonreci_contact}(b-i)), and no contact area during clockwise rotation (see Fig. \ref{fig:res_nonreci_contact}(b-ii)). The high $k_{yy}^-$ is attributed to the contact force being in the opposite direction of the applied force, enhancing the structure's resistance to external forces and resulting in increased stiffness. Additionally, the straight ligament facilitates stretching strain energy when subjected to extension or anti-clockwise rotation. The contact modes of optimal designs for objectives $f_2, f_6$ and $f_8$ are visualized in \ref{apx:contact}, Fig. \ref{fig:apx_nonreci}. 

\subsection{Elastic Asymmetry Optimization}
\label{sec:res_asym}

\begin{figure}[h!]
    \centering
    \includegraphics[width=.9\textwidth]{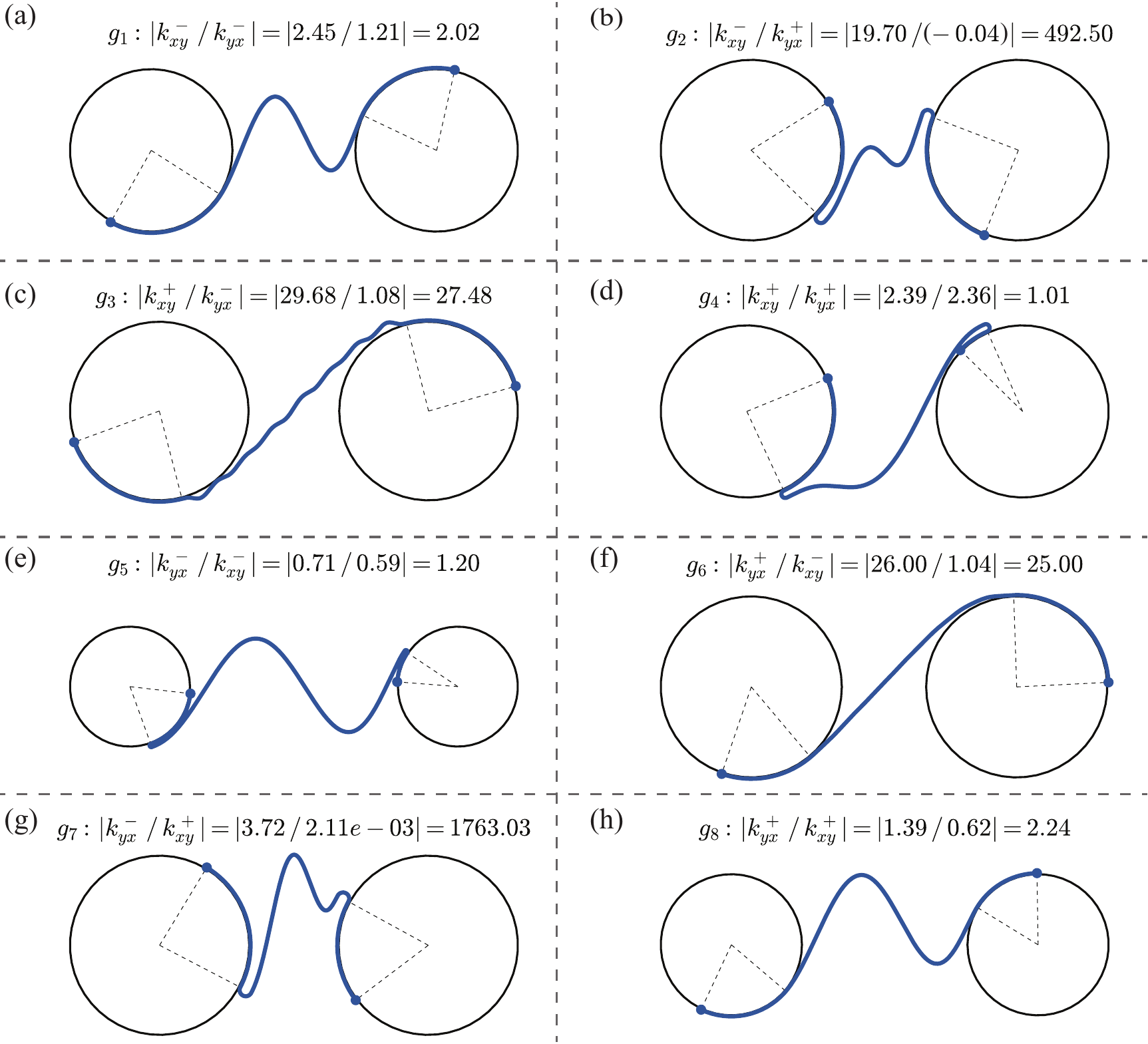}
    \caption{Optimal designs for maximizing eight asymmetry objectives after $10$ iterations of Bayesian Optimization. Each figure shows the optimal design for the (a) objective $g_1$, (b) objective $g_2$, (c) objective $g_3$, (d) objective $g_4$, (e) objective $g_5$, (f) objective $g_6$, (g) objective $g_7$ and (h) objective $g_8$. The title above each structure indicates the corresponding stiffness values.}
    \label{fig:res_asym}
\end{figure}

The optimal designs we discovered for the asymmetry objectives $g_{1:8}$ defined in Section \ref{sec:obj_asym} are presented in Fig. \ref{fig:res_asym}. A summary of the optimal performance for each objective is below

$g_1 = |\frac{k_{xy}^-}{k_{yx}^-}| = |2.45/1.21| = 2.02$, \,
$g_2 = |\frac{k_{xy}^-}{k_{yx}^+}| = |19.70/(-0.04)| = 492.50$, \,
$g_3 = |\frac{k_{xy}^+}{k_{yx}^-}| = |29.68/1.08| = 27.48$, \,

$g_4 = |\frac{k_{xy}^+}{k_{yx}^+}| = |2.39/2.36| = 1.01$, \,
$g_5 = |\frac{k_{yx}^-}{k_{xy}^-}| = |0.71/0.59| = 1.20$, \,
$g_6 = |\frac{k_{yx}^+}{k_{xy}^-}| = |26.00/1.04| = 25.00$, \,

$g_7 = |\frac{k_{yx}^-}{k_{xy}^+}| = |3.72/2.11^{-3}| = 1763.03$, \,
$g_8 = |\frac{k_{yx}^+}{k_{xy}^+}| = |1.39/0.62| = 2.24$.

In contrast to the optimal solutions for non-reciprocity, not all of the optimal designs for asymmetry achieve notably high objective values, as defined by cases where the stiffness of the numerator is larger than the denominator by at least one order of magnitude. Specifically, the optimal designs for $g_2, g_3, g_6$, and $g_7$ achieve high objective values, where the stiffness values in the numerator significantly surpass those in the denominator. These optimal designs exhibit a substantial contact area for one direction but no contact for the other. As a result, the stiffness value for the loading direction where contact is established tend to be much higher, as the contact provides increased resistance to the loads. For a detailed visualization of the contact modes under different load directions, please refer to \ref{apx:contact}, Fig. \ref{fig:apx_asym}.

In contrast, the objective values for $g_1, g_5$, and $g_8$ are relatively small. This is primarily attributed to the challenge of finding designs capable of exhibiting large contact areas in the desired direction while maintaining no contact in the other. In design spaces $1$ and $3$, contact is more likely to occur during the extension and anti-clockwise rotation loads, while no contact is expected for loads from the other two directions. Conversely, in design space $2$, contact is more probable during compression and clockwise rotation, with no contact expected for loads from the other two directions. In design space $4$, contact is likely to occur for loads in all directions. Upon scrutinizing objectives $g_1, g_5$, and $g_8$, it can be found that the required contact modes are not commonly found within these design spaces. For instance, achieving a high value of objective $g_1 = |k_{xy}^-/k_{yx}^-|$ necessitates a large contact area for clockwise rotation loads and no contact for compression, which is not a common scenario within the four design spaces. Nevertheless, we managed to find the optimal design for objective $g_1$ from design space 1. While contact still occurs during compression loading and there is no contact during clockwise rotation, the contact area is very small, mitigating the stiffness value $k_{yx}^-$ during compression. As a result, the final asymmetry results in $k_{xy}^-$ being slightly larger than $k_{yx}^-$. 

\begin{figure}[h!]
    \centering
    \includegraphics[width=.9\textwidth]{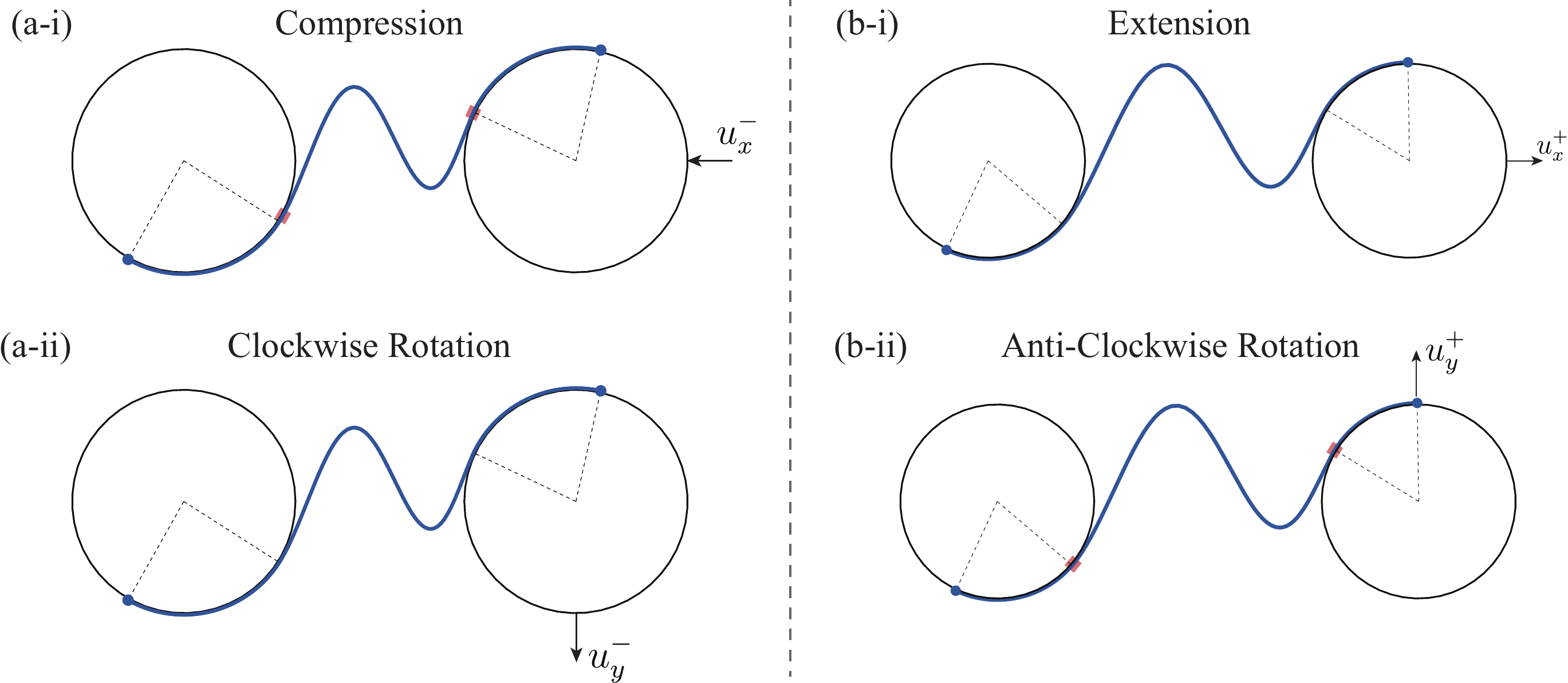}
    \caption{Two examples of contact modes under loads from different directions. The highlighted pink color indicates the contact areas. (a) The contact modes for the optimal design of objective $g_1 = |k_{xy}^-/k_{yx}^-|$. The stiffness value $k_{yx}^-$ is obtained during the (a-i) compression, and $k_{xy}^-$ is obtained during the (a-ii) clockwise rotation. (b) The contact modes for objective $g_8 = |k_{yx}^+/k_{xy}^+|$. The stiffness value $k_{yx}^+$ is obtained during the (b-i) extension, and $k_{xy}^+$ is obtained during the (b-ii) anti-clockwise rotation.}
    \label{fig:res_asym_contact}
\end{figure}

To facilitate the visualization of contact in these optimal structures, we present two examples: the optimal designs for objectives $g_1 = |k_{xy}^-/k_{yx}^-|$ and $g_8 = |k_{yx}^+/k_{xy}^+|$, in Fig. \ref{fig:res_asym_contact}. For both structures, contact modes at the two load directions are different, leading to an asymmetry in the stiffness values. However, the contact area is significantly smaller for the loading direction for which we anticipate the stiffness value to be smaller. 


Finally, the optimal design for $g_4$ yields an objective value of nearly $1$, indicating the absence of designs meeting the criterion where the stiffness value $k_{xy}^+$ exceeds $k_{yx}^+$. To maximize objective $g_4 = |k_{xy}^+/k_{yx}^+|$, an ideal design should feature a substantial contact area under anti-clockwise rotation while exhibiting no contact under extension. As discussed above, such a contact mode is not prevalent within the existing design spaces. The exploration of alternative design spaces capable of accommodating optimal designs meeting the requirements of objective $g_4$ remains a topic for future investigation.

\subsection{Multi-Objective Optimization}
\label{sec:res_multi}
In this Section, our goal is to identify optimal designs that exhibit both non-reciprocity and asymmetry, as defined in Section \ref{sec:obj_multi}. Therefore, we aim to optimize two objectives, selecting one from $f_{1:8}$, and another from $g_{1:8}$. Based on the results of single objective optimization in Sections \ref{sec:res_nonreci} and \ref{sec:res_asym}, we observed that the optimal designs for objectives $f_1, f_3, f_5, f_7, g_2$, and $g_7$ exhibit similarities, which we designate as Group 1 since the objectives are not contradictory. Similarly, the optimal designs for objectives $f_2, f_4, f_6, f_8, g_3$, and $g_6$ share similarities and are classified as Group 2. It's important to note that the objectives in Group 1 and Group 2 are contradictory because their optimal designs for single objective optimization exhibit opposite characteristics. Additionally, we assign objectives $g_1, g_5$, and $g_8$ as Group 3 due to their relatively small optimal values discovered in Section \ref{sec:res_asym}. Objective $g_4$ does not have an optimal design in the single objective optimization, so it is not considered in the multi-objective optimization scenario. The three groups of objectives are summarized below.

\begin{itemize}
    \item Group 1 (non-contradictory): $f_1, \, f_3, \, f_5, \, f_7, \, g_2, \, g_7$
    \item Group 2 (non-contradictory): $f_2, \, f_4, \, f_6, \, f_8, \, g_3, \, g_6$
    \item Group 3 (challenging): $g_1, \, g_5, \, g_8$
\end{itemize}

Based on the different groups established above for the objectives, we explore three different cases for multi-objective optimization. In the first case, we consider non-contradictory objectives in the same group, where both objectives are chosen from Group 1 or both chosen from Group 2. As a representative case, we select $f_1$ and $g_2$ from Group 1. In the second case, we choose two contradictory objectives, one from Group 1 and another from Group 2. The representative case we use is $f_1$ and $g_3$. In the third case, we choose one challenging objective from Group 3 and one relatively feasible objective from Group 1 or Group 2. The representative case we use is $f_1$ and $g_1$. The distribution of the objectives for the three cases can be found in Fig. \ref{fig:apx_case1} - \ref{fig:apx_case3}. Below is a summary of the three representative multi-objective optimization cases:

\begin{itemize}
    \item Multi-Objective 1 (non-contradictory): $f_1$ and $g_2$
    \item Multi-Objective 2 (contradictory): $f_1$ and $g_3$
    \item Multi-Objective 3 (challenging): $f_1$ and $g_1$
\end{itemize}

\begin{figure}[h!]
    \centering
    \includegraphics[width=1\textwidth]{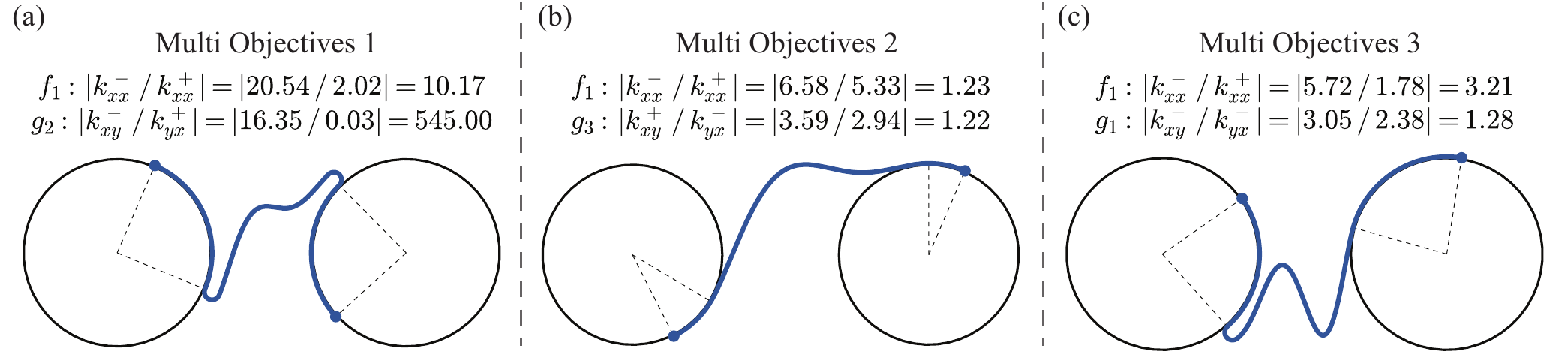}
    \caption{The Pareto front and the corresponding designs for optimizing (a) Multi-Objective 1: $f_1$ and $g_1$, (b) Multi-Objective 2: $f_1$ and $g_3$, (c) Multi-Objective 3: $f_1$ and $g_1$. The title above each structure indicates the corresponding objective values for the structure.}

    \label{fig:res_multi}
\end{figure}

After $10$ iterations of Bayesion Optimization, we identified three Pareto optima for Multi-Objective 1, one Pareto optimum for Multi-Objective 2 and four for Multi-Objective 3. The Pareto front of each multi-objective and its corresponding designs can be found in \ref{apx:multi_obj}. For each Multi-Objective, we selected the Pareto optima with the largest modulus (i.e. square root of the sum of all the objective values) and presented them in Fig. \ref{fig:res_multi}. In Multi-Objective 1, where the objectives $f_1$ and $g_2$ are not contradictory, both objectives achieved relatively high values with magnitudes larger than ten, as depicted in Fig. \ref{fig:res_multi}(a). However, Multi-Objective 2 presents a challenge as it involves two contradictory objectives, $f_1$ and $g_3$, where increasing one objective results in a decrease in the other, as illustrated in Fig. \ref{fig:apx_case2}. The optimal design discovered by ML methods shows a trade-off between asymmetry and non-reciprocal properties, albeit with relatively smaller values for both objectives, as shown in Fig. \ref{fig:res_multi}(b). Similarly, Multi-Objective 3 includes a challenging objective, $g_1$, and an easily attainable objective, $f_1$. As shown in Fig. \ref{fig:apx_case3}, the high value of $g_1$ is observed only when $f_1$ is small, indicating a contradiction between the two objectives. The optimal design after balancing these conflicting objectives is presented in Fig. \ref{fig:res_multi}(c). Notably, the optimal designs for the contradictory multi-objectives 2 and 3 are both found in Design Space 4 (refer to Fig. \ref{fig:ds}), while all the optimal design are obtained from Design Space 1-3 when we optimize each single objective. This highlights that optimizing a single objective does not necessarily ensure high performance across all dimensions. Leveraging the Pareto Front and Bayesian Optimization facilitates the discovery of designs that exhibit multiple desirable properties simultaneously.

\begin{figure}[h!]
    \centering
    \includegraphics[width=.9\textwidth]{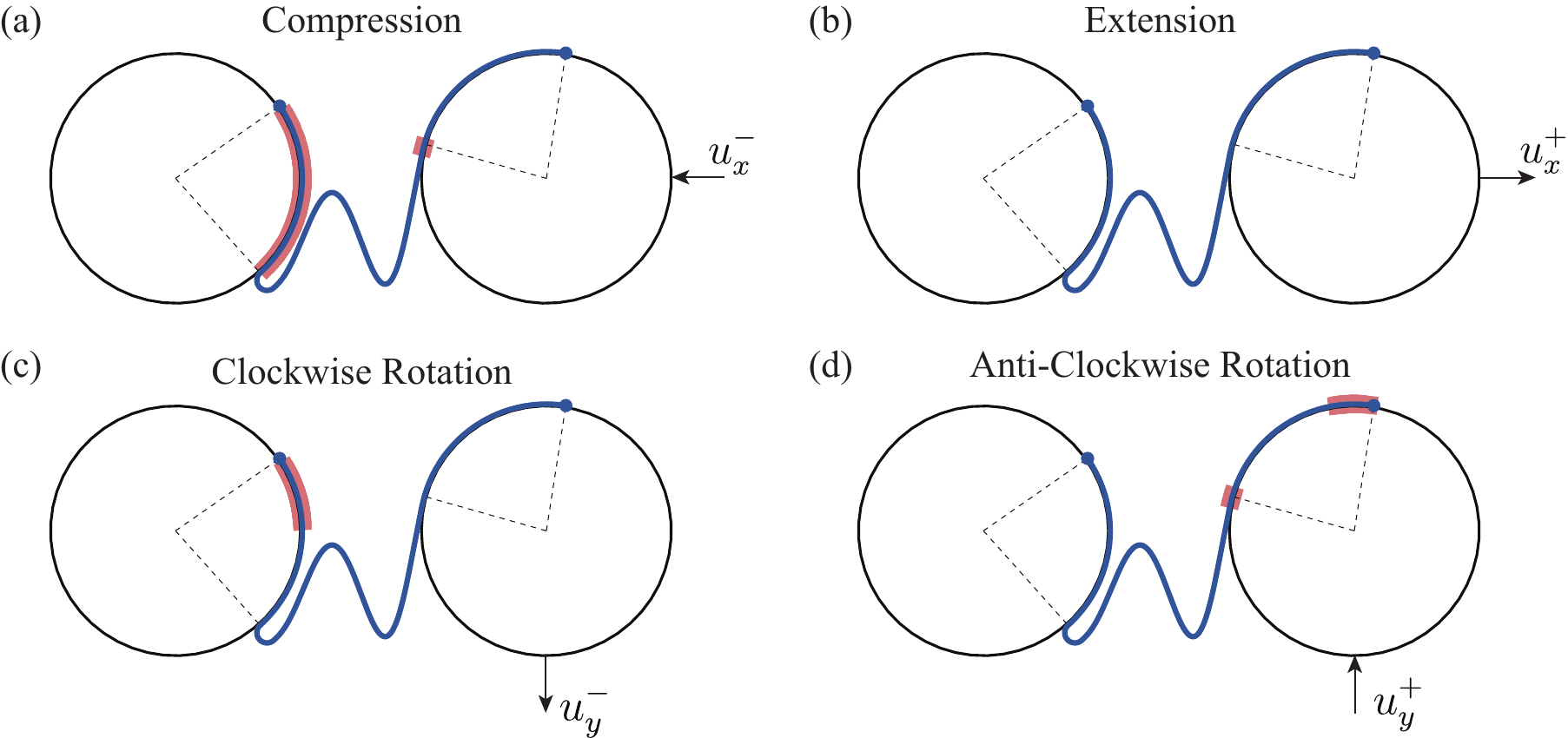}
    \caption{The contact modes for the optimal design shown in Fig. \ref{fig:res_multi}(c). The highlighted pink color indicates
the different contact areas between the rigid circle and the elastic ligament under (a) Compression, (b) Extension, (c) Clockwise Rotation, and (d) Anti-Clockwise Rotation. The structure satisfies the most objectives in terms of achieving both non-reciprocity and asymmetry that are summarized in Section \ref{sec:res_multi}.}
    \label{fig:res_multi_contact}
\end{figure}

Upon scrutinizing the contact modes of these optimal designs, we observed a notable distinction in the optimal design depicted in Fig. \ref{fig:res_multi}(c) for Multi-Objective 3 compared to the others. Specifically, while the majority of chiral structures typically exhibit only two distinct contact states under loads from four different directions, as illustrated in the example case in Fig. \ref{fig:obj}, the structure in Fig. \ref{fig:res_multi}(c) displays four different types of contact states under loads from four different directions. The four distinct contact statuses are illustrated in Fig. \ref{fig:res_multi_contact}, where the highlighted pink areas denote the contact regions under each load. Notably, the contact areas vary depending on the applied loads. Since each contact mode corresponds to beam models with different boundary conditions, as discussed in Section \ref{sec:contact}, the stiffness values vary in all directions, resulting in $[k_{xx}^-,\,  k_{xy}^-,\,  k_{yx}^-,\,  k_{yy}^-, \, k_{xx}^+,\,  k_{xy}^+, \, k_{yx}^+, \, k_{yy}^+] = [5.72, \, 3.05, \, 2.38, \, 3.51, \, 1.78, \, 0.42, \,  1.13,\,  2.11]$. Consequently, this structure can satisfy the most objectives in terms of achieving both non-reciprocity and asymmetry, summarized below:

$f_1 = |\frac{k_{xx}^-}{k_{xx}^+}| = {5.72}/{1.78} = 3.21$, \, 
$f_3 = |\frac{k_{xy}^-}{k_{xy}^+}| = {3.05}/{0.42} = 7.26$, \,
$f_5 = |\frac{k_{yx}^-}{k_{yx}^+}| = {2.38}/{1.13}= 2.10$, 

$f_7 = |\frac{k_{yy}^-}{k_{yy}^+}| = {3.51}/{2.11} = 1.66$, \,
$g_1 = |\frac{k_{xy}^-}{k_{yx}^-}| = {3.05}/{2.38} = 1.28$, \,
$g_2 = |\frac{k_{xy}^-}{k_{yx}^+}| = {3.05}/{1.13} = 2.69$, 

$g_7 = |\frac{k_{yx}^-}{k_{xy}^+}| = {2.38}/{0.42} = 5.66$, \,
$g_8 = |\frac{k_{yx}^+}{k_{xy}^+}| = {1.13}/{0.42} = 2.69$.

\section{Conclusion}
In this paper, we utilized machine learning (ML) techniques in conjunction with Finite Element Simulation to engineer chiral metamaterials exhibiting significant directional non-reciprocity and stiffness asymmetry. We elucidated the mechanisms underlying non-reciprocity and asymmetry by qualitatively analyzing the contact behavior under various loads through equivalent beam models. Additionally, through quantitatively analysis of strain energy changes during chiral metamaterial deformation, we uncovered that stretching deformation of the elastic ligament in the chiral structures leads to higher stiffness, while the bending structures result in lower stiffness values. This insight explains the heuristic behind programming the chiral metamaterial geometry to achieve desired non-reciprocity and asymmetry properties.

Our study encompassed optimization of eight non-reciprocity objectives, eight asymmetry objectives, and three multi-objectives incorporating both non-reciprocity and asymmetry. To formalize the problem under the framework of ML, we defined the design freedom of chiral structures and characterized each structure by its geometric features. We proposed four different design spaces, with each being governed by a different contact mechanism. To reduce the computational cost, we implemented Bayesian Optimization algorithm facilitated with ensemble learning to efficiently search the optimal structures for each objective by balancing between the exploration and exploitation. Particularly, we achieve the multi-objective optimization by searching the Pareto Front of the performance spaces and obtained the corresponding optimal structures that can exhibit both non-reciprocity and asymmetry. Following the completion of the optimization, we presented the discovered optimal designs for each single objective and analyzed their similarities and disparities, highlighting the role of contact in enhancing chiral structure stiffness. Leveraging insights from single-objective optimization, we focused on three distinct types of multi-objectives, including both contradictory and non-contradictory combination of multi-objectives. Among the discovered optimal designs for multi-objectives, we found a structure that demonstrates the maximum diversity in non-reciprocity and asymmetry, which we attributed to the various contact behavior of the structure under different loads. 

In contrast to the majority of the current ML and mechanics literature focusing on testing the applicability of machine learning (ML) methods in solving conventional mechanics problems, where their effectiveness can be validated against known solutions~\citep{brunton2016discovering, chen2021learning,jiao2024machine}, our study employed ML to a novel domain with limited prior information. With the aid of ML, we uncovered a compelling relationship between contact behaviors and the non-reciprocity and asymmetry of chiral metamaterials through qualitative, quantitative, and case-based analyses. A noteworthy observation from our results is the presence of extreme non-reciprocity and asymmetry in the discovered optimal designs. For instance, as illustrated in Fig. \ref{fig:res_asym}(g), the optimal designs for the objective $g_7$ have a stiffness $k_{yx}^{-}$ that is $1763$ times larger than $k_{xy}^{+}$. The extreme and diverse non-reciprocity and asymmetry observed in our optimized chiral structures pave the way for metamaterials capable of novel wave propagation characteristics, including unidirectional wave propagation and non-Hermitian wave phenonena~\citep{wangJMPS2024}, which offers promising prospects for applications in non-reciprocal wave propagation. 



While we have optimized the chiral metamaterial to achieve various level of non-reciprocity and asymmetry, we noticed that some of the asymmetry objectives did not achieve notably high objective values, and the objective $g_4$ resulted in no qualified optimal design within the current scope of design spaces. Additionally, although the optimal design for the multi-objective optimization displayed diverse non-reciprocity and asymmetry characteristics, as depicted in Fig. \ref{fig:res_multi_contact}, the absolute magnitudes of most objective values remained relatively small. We attribute this to the current confined design spaces, i.e., the rigid body of the structure can only be circles, and the shape of partially connected elastic ligament are well-defined by certain base functions, as detailed in Section \ref{sec:ds}. Looking ahead, we aim to broaden the design scope to explore a wider range of possibilities and search for designs capable of exhibiting extreme multi-objective performance. Furthermore, leveraging deep generative models such as GANs (Generative Adversarial Networks) ~\citep{wang2022ih,kobeissi2022enhancing}, Transformers ~\citep{buehler2024mechgpt, buehler2023generative}, and Diffusion Models ~\citep{ni2024forcegen,luu2023generative}, we anticipate for advanced AI-designed structures with non-conventional properties, providing insights not only into the field of odd elastic materials but also into other emerging novel domains.

\section{Additional Information} 
\label{sec:additional_info}

The data supporting this study and the code to optimize chiral metamaterial designs and reproduce simulation on ABAQUS are openly available from \url{https://github.com/lingxiaoyuan/chiral}.

\section{Acknowledgements} 
All authors gratefully acknowledge the support of the College of Engineering and Department of Mechanical Engineering at Boston University, the Hariri Institute for Computing Junior Faculty Fellowship, the David R. Dalton Career Development Professorship, as well as the helpful discussions with Peerasait (Jeffrey) Prachaseree and Yan Yang.

\appendix
\section{Finite Element Simulation}
\label{apx:fea}
\begin{figure}[h!]
    \centering
    \includegraphics[width=.6\textwidth]{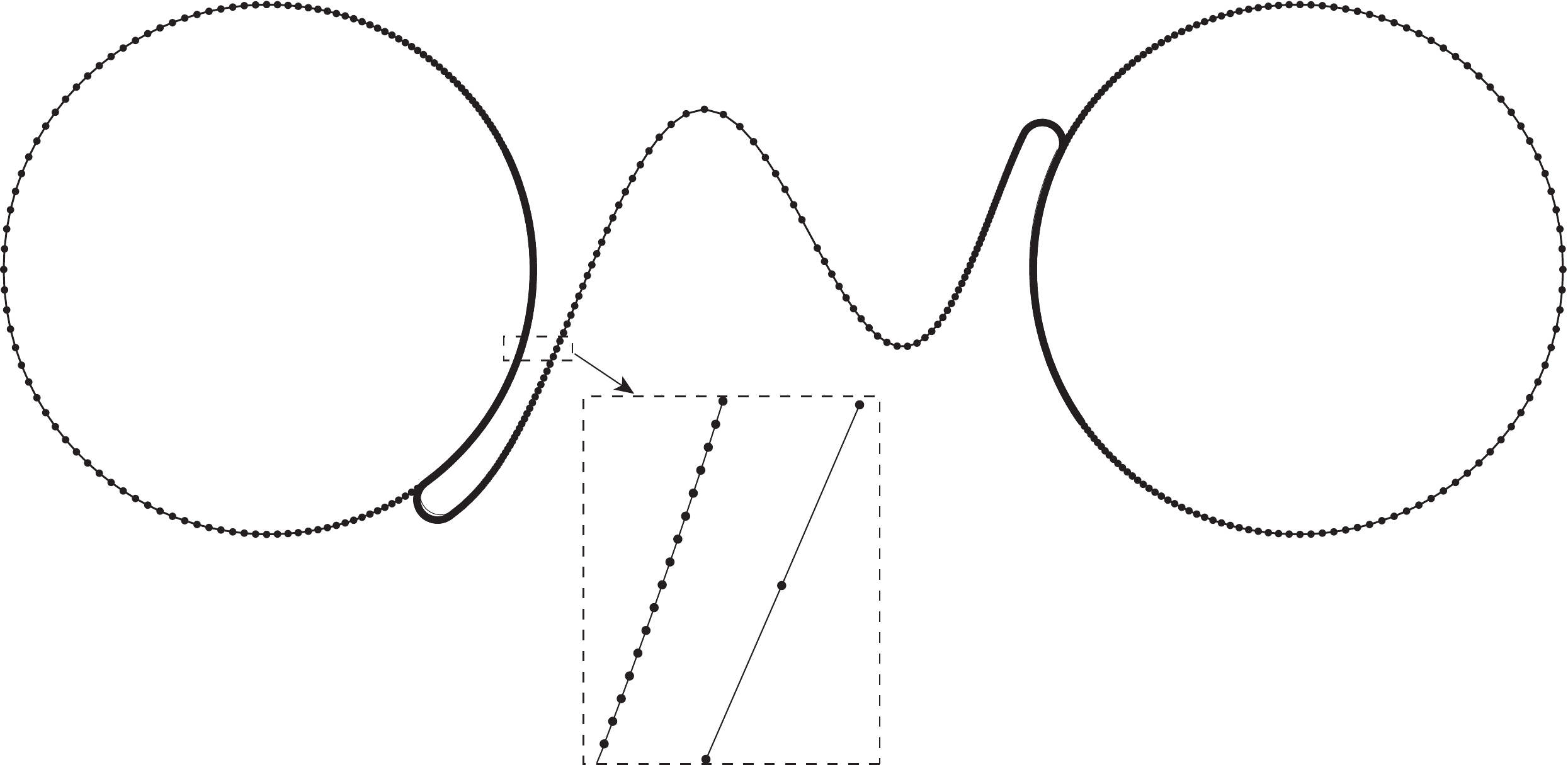} 
    \caption{Visualization of the mesh on a representative chiral structure. The finest mesh size is $0.02$ and the coarsest mesh size is $0.4$. Graded meshing was performed using the commercial software ABAQUS. }
    \label{fig:mesh}
\end{figure}

\begin{figure}[h!]
    \centering
    \includegraphics[width=1.0\textwidth]{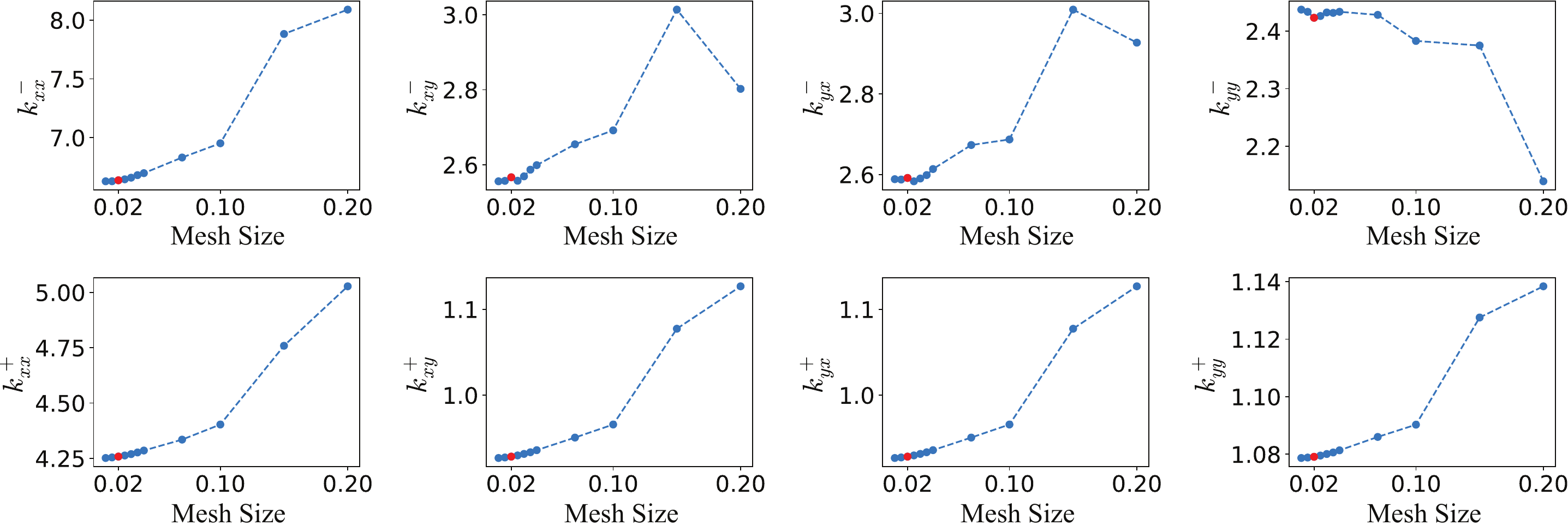} 
    \caption{The relationship between calculated stiffness values and the selected finest mesh size for the chiral structure depicted in Fig. \ref{fig:mesh}. The final adopted finest mesh size is $0.02$, highlighted in red.}
    \label{fig:sensitivity}
\end{figure}

The properties of the chiral metamaterial were acquired through the Finite Element Method (FEM). In this section, we provide detailed information on the FEM settings. The two rigid circles are set as discrete rigid surfaces, while the elastic ligament of each chiral structure is modeled as linear elastic Bernoulli–Euler beam. The cross section of the ligament is rectangular with a fixed width of $30$ and a fixed thickness $1.5$. The ligament material is linear elastic with a fixed Elastic modulus $E = 70$ and Poisson ratio $\nu =0.3$. The left circle is fixed, while the right rigid circle is only allowed to move under the applied displacement load with same magnitude $0.08$. Both the circle and the ligament are meshed in a graded fashion, where the area closer to the connecting region has a finer mesh compared to areas farther away. The finest mesh size is set as $0.02$ and the coarsest mesh size is $20$ times the finest mesh size. On average, there are approximately $270$ mesh elements on each of the rigid circles and approximately $714$ mesh elements on the elastic ligament, as calculated from $12$ randomly generated chiral structure finite element models. Fig. \ref{fig:mesh}, exported from the sofware ABAQUS, illustrates the distribution of mesh size on the rigid circles and the elastic ligament for a representative example. Each circle has $254$ R2D2 (two node 2D linear discrete rigid element) elements  and the ligament has $920$ B23 (two node cubic Euler-Bernoulli beam element) elements \citep{AbaqusAnalysis}. The mesh size at the ligament-circle contact regions is smaller than other regions. The finest mesh size $0.02$ was determined by conducting mesh sensitivity analysis. Fig. \ref{fig:sensitivity} demonstrates how stiffness values change depending on the mesh size. Choosing an appropriate mesh size involves balancing computational cost and FEM accuracy. It is worth noting that excessively fine mesh sizes, such as $0.01$, can lead to more simulation failures due to convergence difficulties. Therefore, a mesh size of $0.02$, where simulation results become converged and nearly independent of mesh size, was adopted for all chiral metamaterial simulations in this paper. Nonetheless, we note that a few simulation cases still fail to converge for certain chiral structures with a mesh size of $0.02$. These failed examples were disregarded during the Bayesian Optimization data acquisition process. For more details on the finite element modeling, readers can refer to the open-source code we published on GitHub, as detailed in Section \ref{sec:additional_info}.

\section{Contact modes for optimal designs}
\label{apx:contact}

\begin{figure}[p]
    \centering
    \includegraphics[width=.7\textwidth]{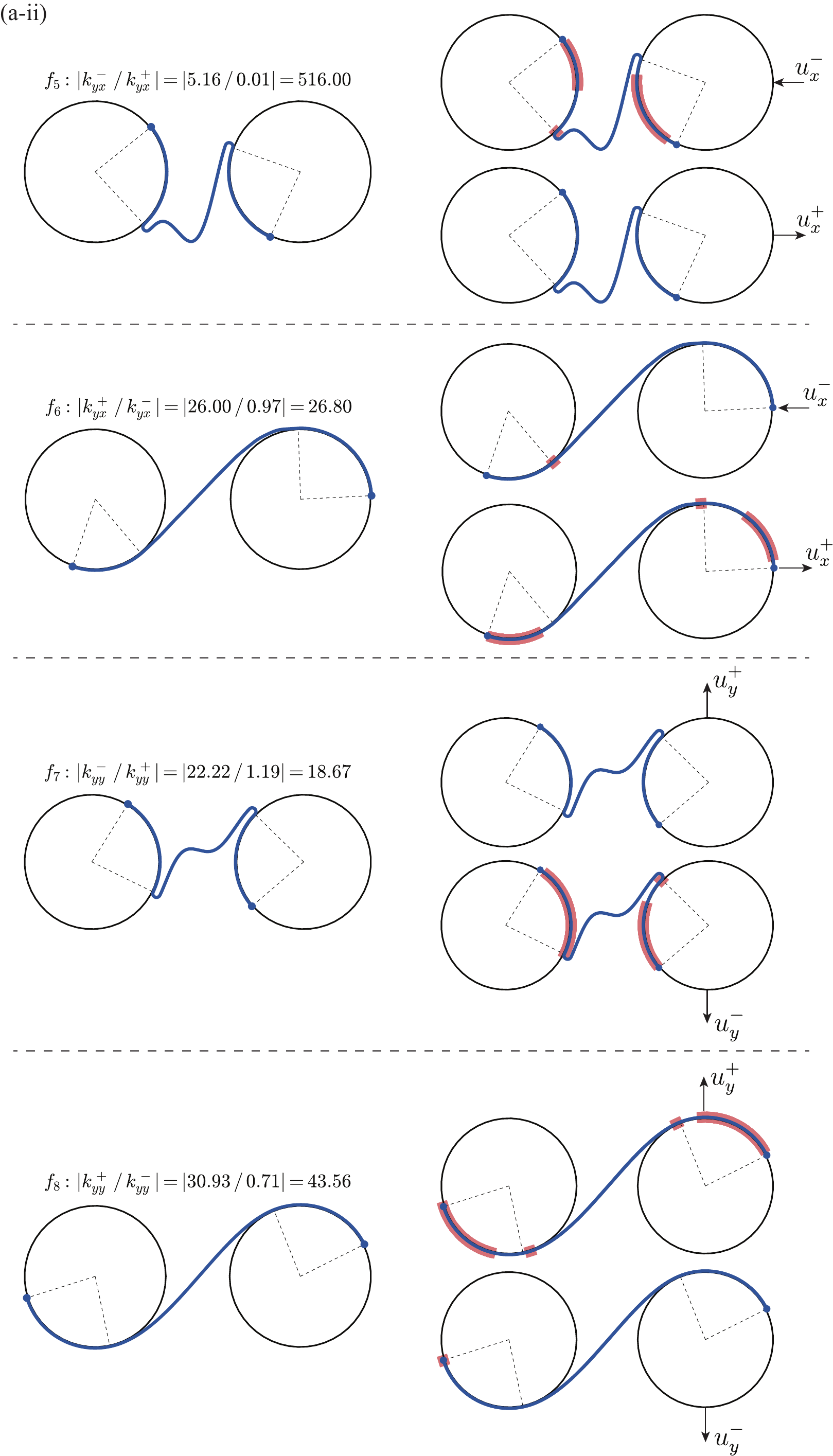}
    \caption{The contact modes for the optimal designs of the eight nonreciprocity objectives (a-i) $f_1, f_2, f_3, f_4$ and (a-ii) $f_5, f_6, f_7, f_8$. For each objective, the optimal design and the contact modes during the loading in which the stiffness values are obtained are depicted.}
    \label{fig:apx_nonreci}. 
\end{figure}

\begin{figure}[p]
    \centering
    \includegraphics[width=.7\textwidth]{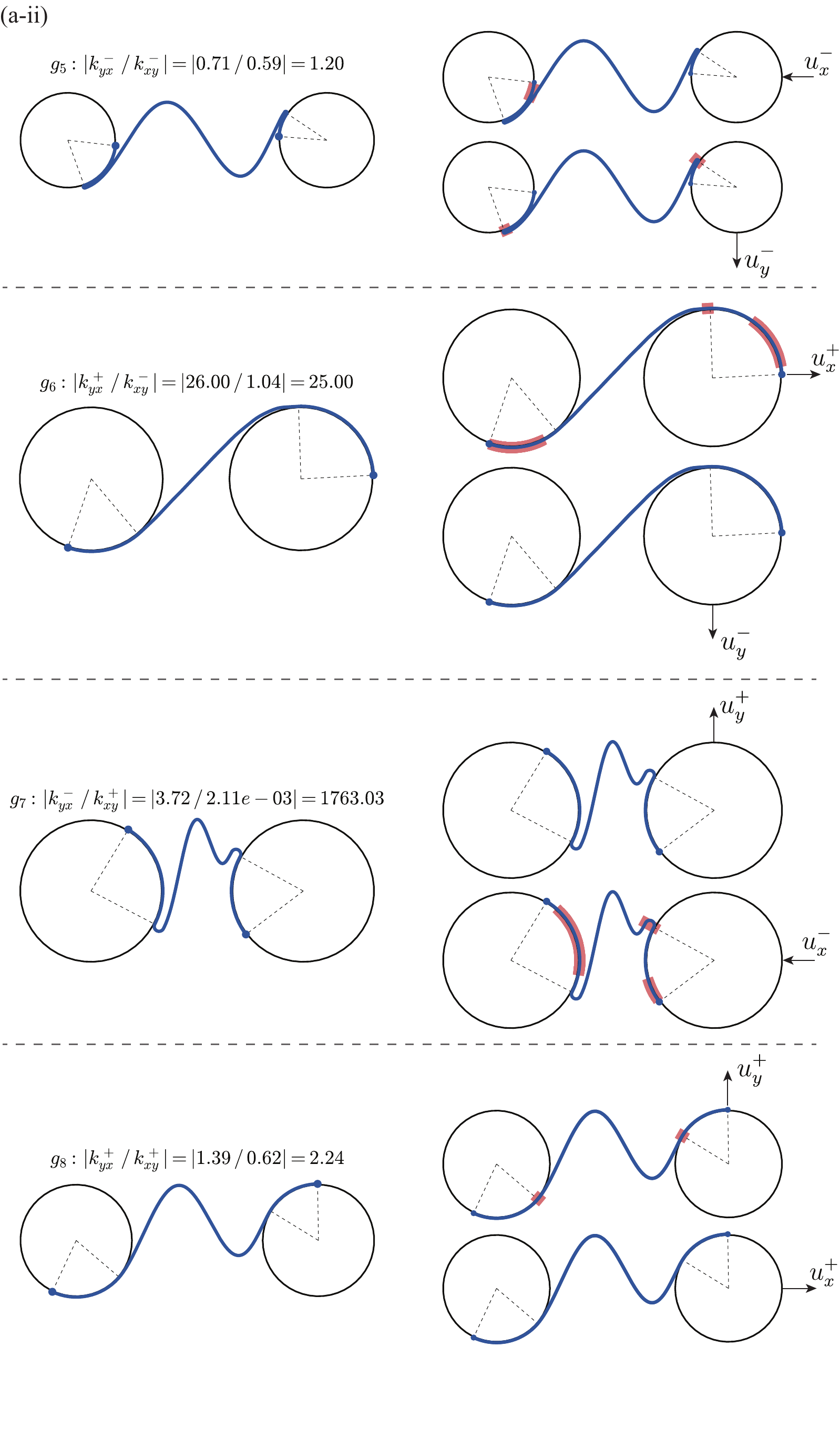}
    \caption{The contact modes for the optimal designs of the eight asymmetry objectives (a-i) $g_1, g_2, g_3, g_4$ and (a-ii) $g_5, g_6, g_7, g_8$. For each objective, the optimal design and the contact modes during the loading in which the stiffness values are obtained are depicted.}
    \label{fig:apx_asym}. 
\end{figure}

In Section \ref{sec:res_nonreci} and Section \ref{sec:res_asym}, we presented the optimal designs for the eight objectives of non-reciprocity and eight objectives of asymmetry. In this Section, we presented the contact status for each design under different loads. Specifically, Fig. \ref{fig:apx_nonreci} shows the optimal chiral structures and the corresponding contact status for objectives $f_1$ to $f_8$, Fig. \ref{fig:apx_asym} shows the optimal chiral structures and the corresponding contact status for objectives $g_1$ to $g_8$.

\begin{figure}[p]
    \centering
    \includegraphics[width=.7\textwidth]{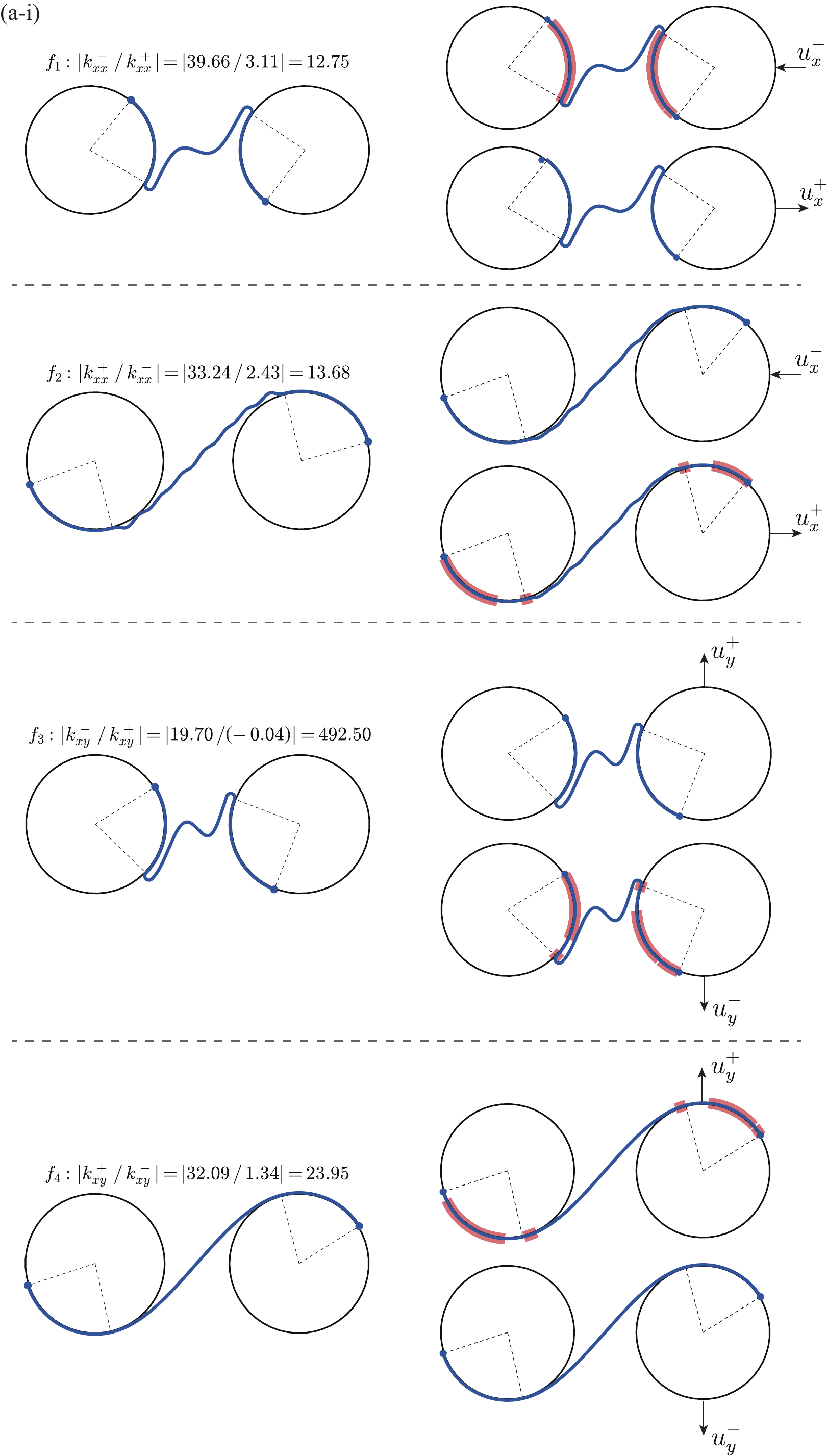} 
\end{figure}

\begin{figure}[p]
    \centering
    \includegraphics[width=.7\textwidth]{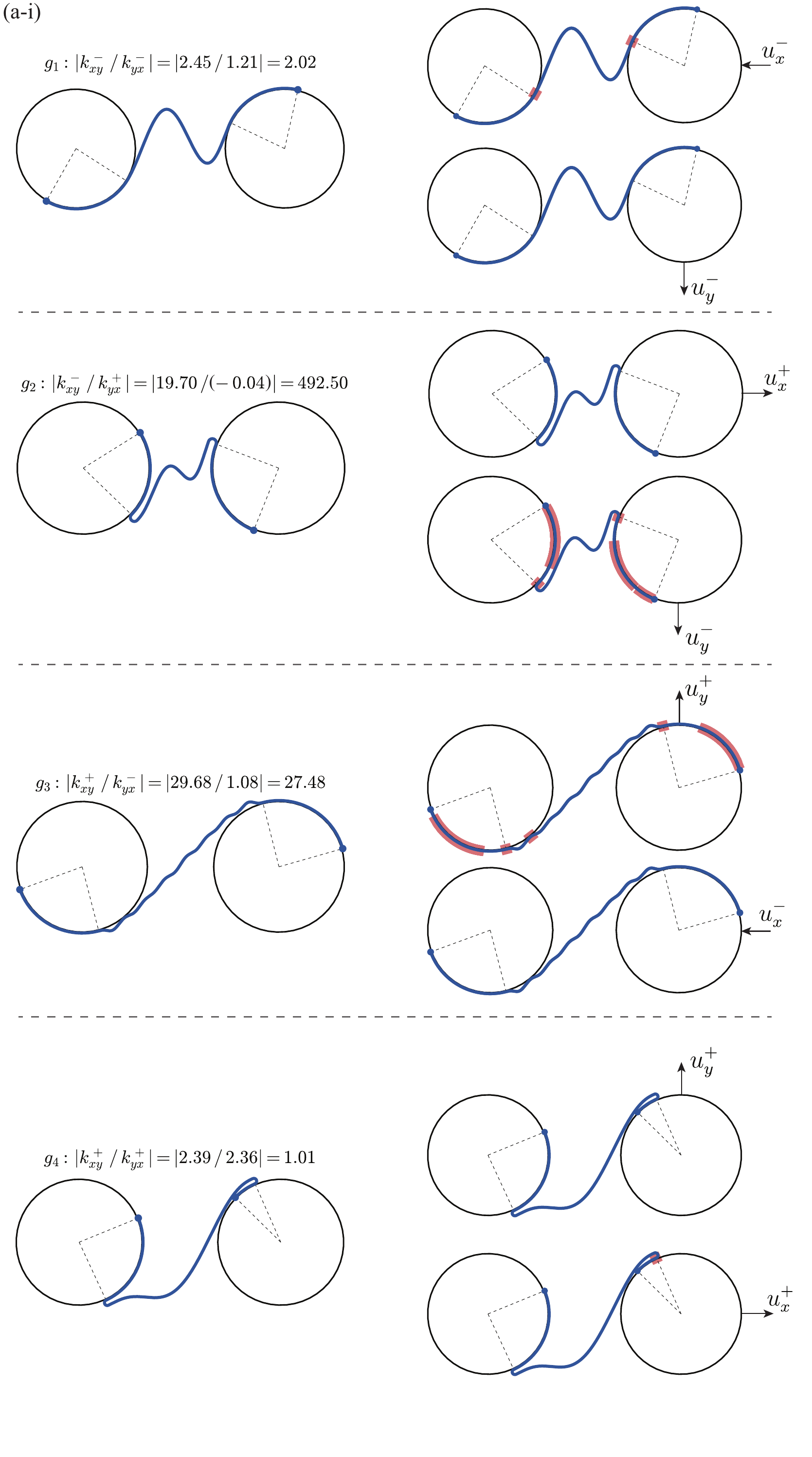} 
\end{figure}

\section{Data distribution for multi-objectives}

\begin{figure}[h!]
    \centering
    \includegraphics[width=1.\textwidth]{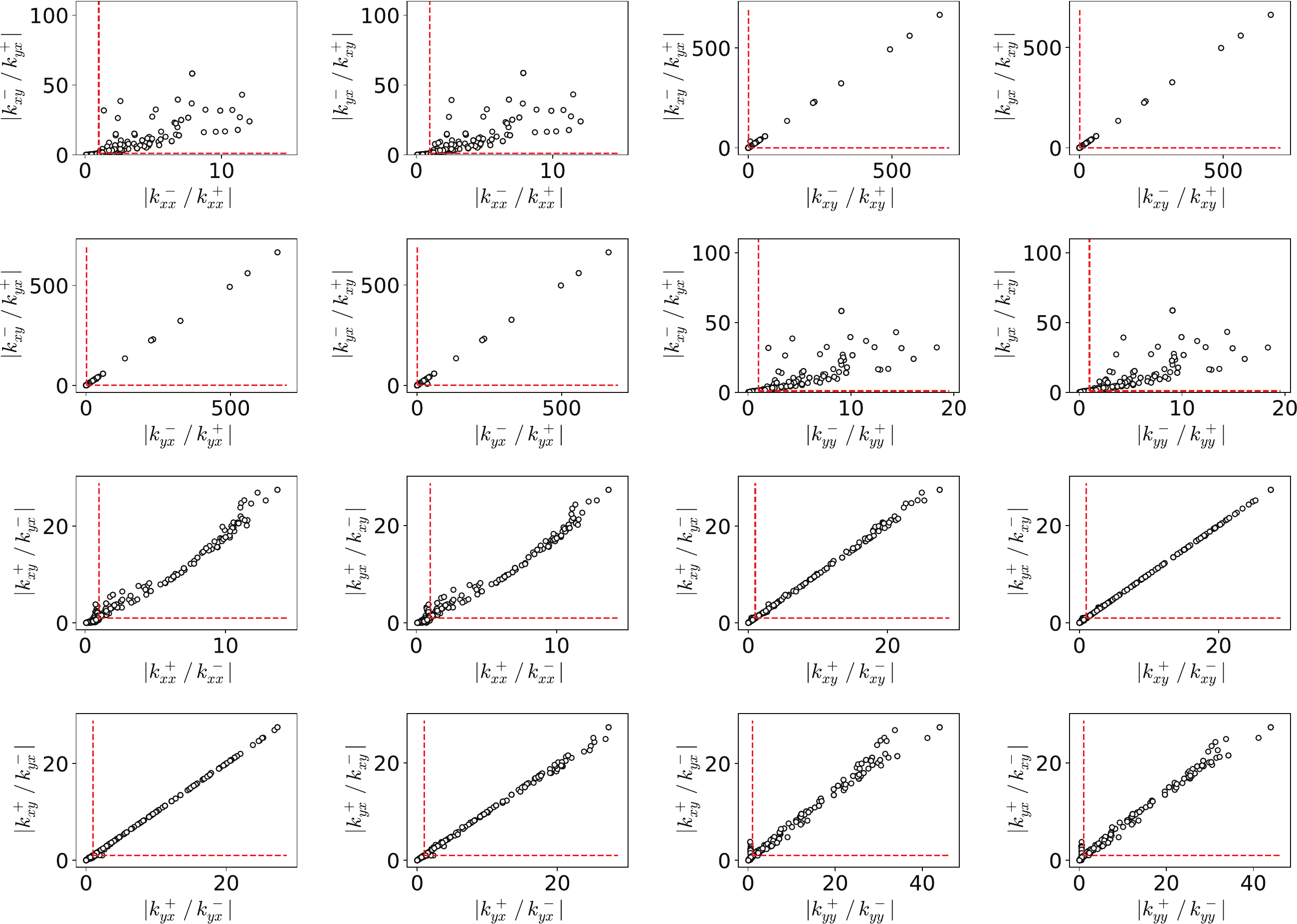}
    \caption{The data distribution of $16$ pairs of non-contradictory multi-objectives for non-reciprocity and asymmetry multi-objectives. The red dashed lines indicate thresholds of $x=1$ and $y=1$ for valid designs. The $x$-axis denotes one non-reciprocity objective and the $y$-axis denotes one asymmetry objective, as defined in Sections \ref{sec:obj_nonreci} and Section \ref{sec:obj_asym} respectively.}
    \label{fig:apx_case1}
\end{figure}

\begin{figure}[h!]
    \centering
    \includegraphics[width=1.\textwidth]{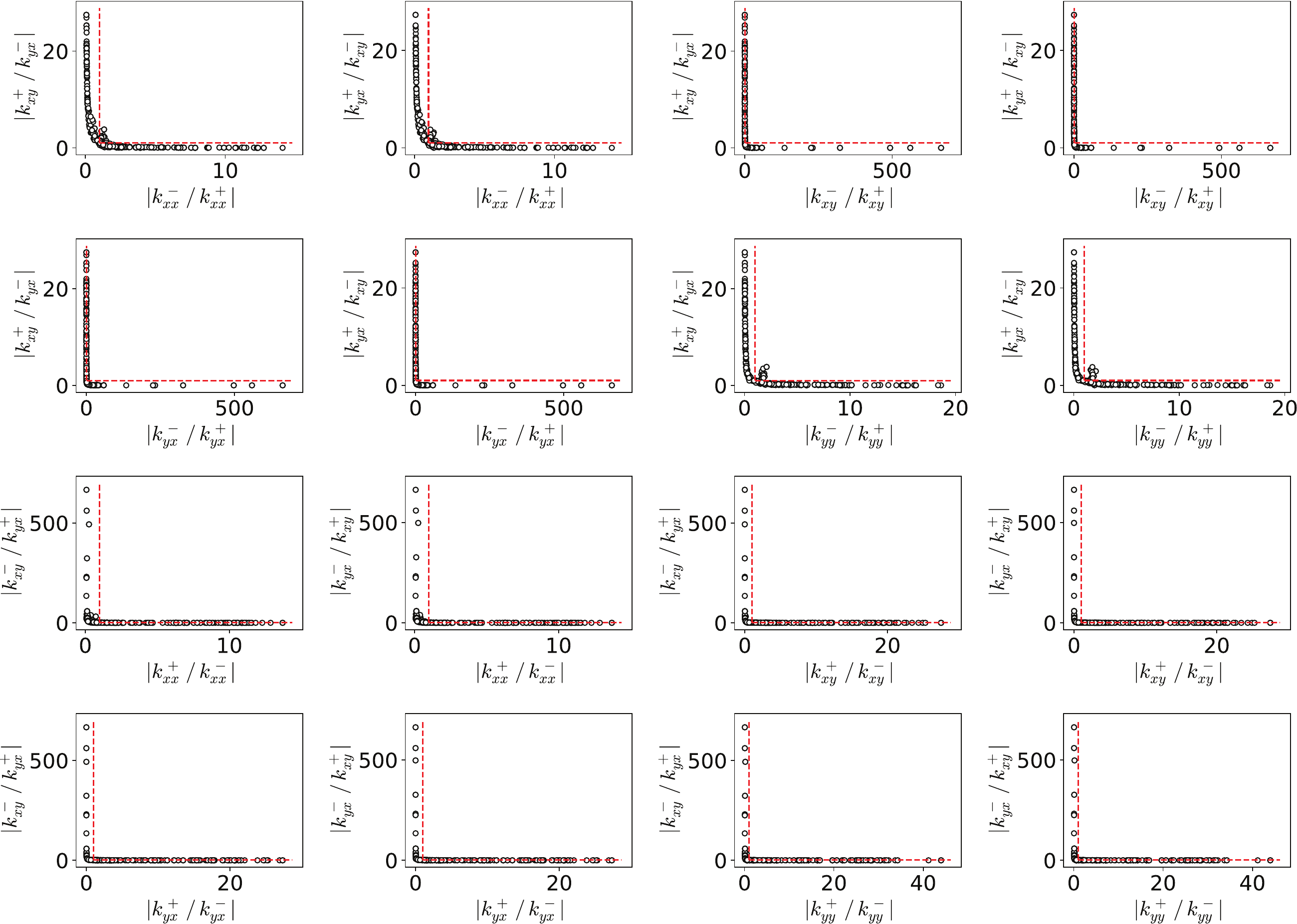}
    \caption{The data distribution of $16$ pairs of contradictory multi-objectives for non-reciprocity and asymmetry multi-objectives. The red dashed lines indicate thresholds of $x=1$ and $y=1$ for valid designs. The $x$-axis denotes one non-reciprocity objective and the $y$-axis denotes one asymmetry objective, as defined in Sections \ref{sec:obj_nonreci} and Section \ref{sec:obj_asym} respectively.}
    \label{fig:apx_case2}
\end{figure}

\begin{figure}[h!]
    \centering
    \includegraphics[width=.9\textwidth]{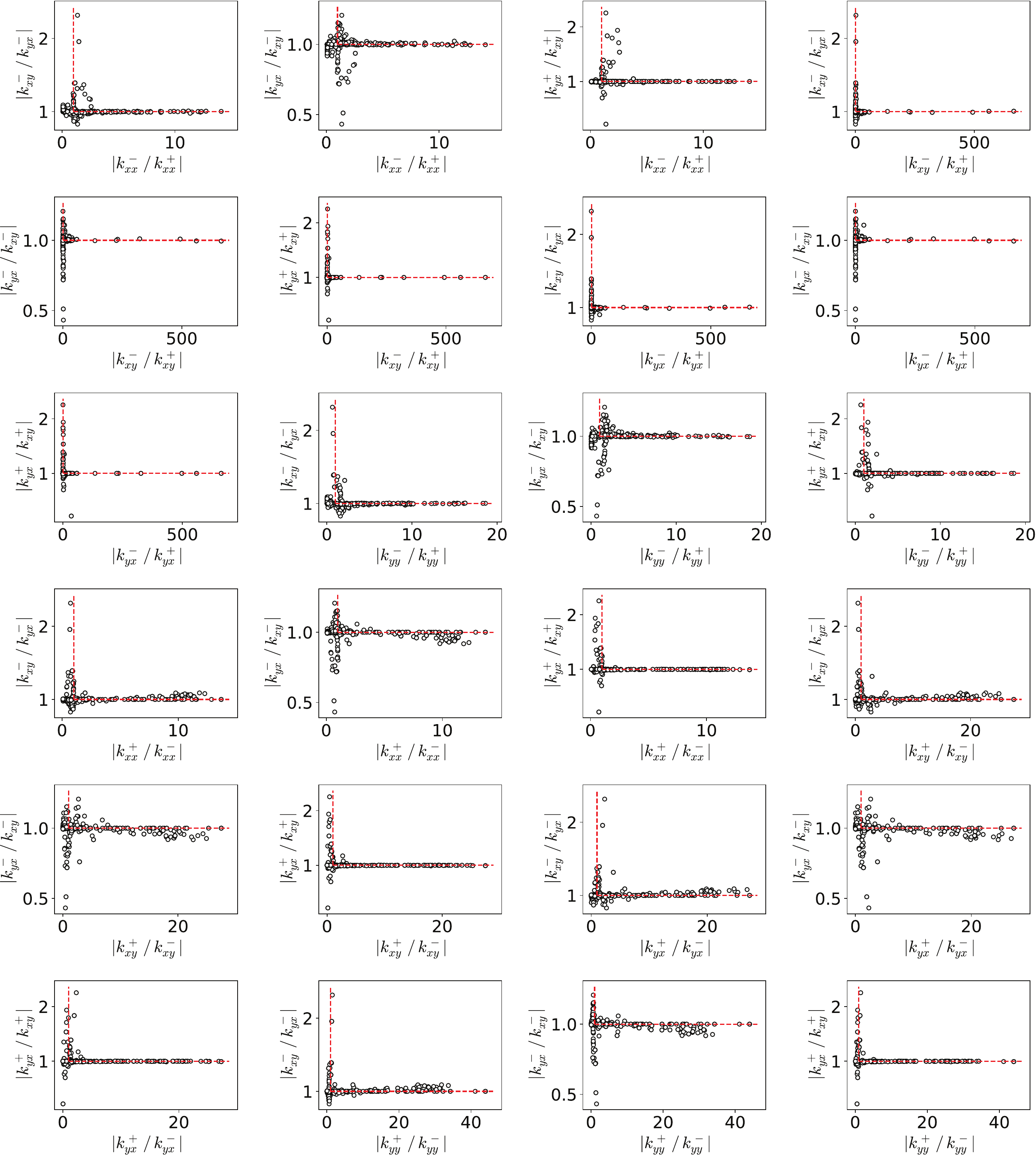}
    \caption{The data distribution of $24$ pairs of challenging multi-objectives for non-reciprocity and asymmetry multi-objectives. The red dashed lines indicate thresholds of $x=1$ and $y=1$ for valid designs. The $x$-axis denotes one non-reciprocity objective and the $y$-axis denotes one asymmetry objective, which is the objective that is challenging to optimize. The definition of non-reciprocity and asymmetry can be found in Sections \ref{sec:obj_nonreci} and Section \ref{sec:obj_asym}.}
    \label{fig:apx_case3}
\end{figure}

We illustrated the trade-off between asymmetry and non-reciprocity in multi-objective optimization in Section \ref{sec:obj_multi}, depicted in Fig. \ref{fig:pareto}, where we aim to optimize two objectives with values exceeding $1$ while maintaining good performance for both. With eight non-reciprocity and eight asymmetry objectives defined in Sections \ref{sec:obj_nonreci} and Section \ref{sec:obj_asym} respectively, there are $64$ comprehensive combinations of non-reciprocity and asymmetry multi-objective optimization. In Section \ref{sec:res_multi}, we categorized multi-objective optimization into three cases: non-contradictory, contradictory, and challenging. To provide insight into this categorization, we randomly sampled $500$ chiral structures obtained in the single-objective optimization process and plotted their distribution of non-reciprocity and asymmetry properties in Fig. \ref{fig:apx_case1}-\ref{fig:apx_case3}, where the red dashed lines represent thresholds of $x=1$ and $y=1$ for valid designs. Any design having a non-reciprocity or asymmetry property below $1$ was not considered further. Fig. \ref{fig:apx_case1} demonstrates $16$ pairs of non-reciprocity and asymmetry where the properties are not contradictory, allowing for simultaneous improvement of both. Conversely, Fig. \ref{fig:apx_case2} exhibits $16$ contradictory pairs, where high performance in one property corresponds to low performance in the other. Fig. \ref{fig:apx_case3} illustrates the data distribution of $24$ pairs of multi-objectives, introducing a challenging asymmetry property that makes it difficult to achieve good performance in non-reciprocity simultaneously. Notably, the non-reciprocity of all materials generally falls below 3. Note that the the total number of pairs displayed in Fig. \ref{fig:apx_case1}-\ref{fig:apx_case3} is less than $64$ due to the exclusion of objective $g_4 = k_{xy}^+/k_{yx}^+$, for which no optimal design was found during the single-objective optimization process, as discussed in Section \ref{sec:res_asym} and Section \ref{sec:res_multi}.

\section{Pareto Front for multi-objective optimization}
\label{apx:multi_obj}

\begin{figure}[h!]
    \centering
    \includegraphics[width=1\textwidth]{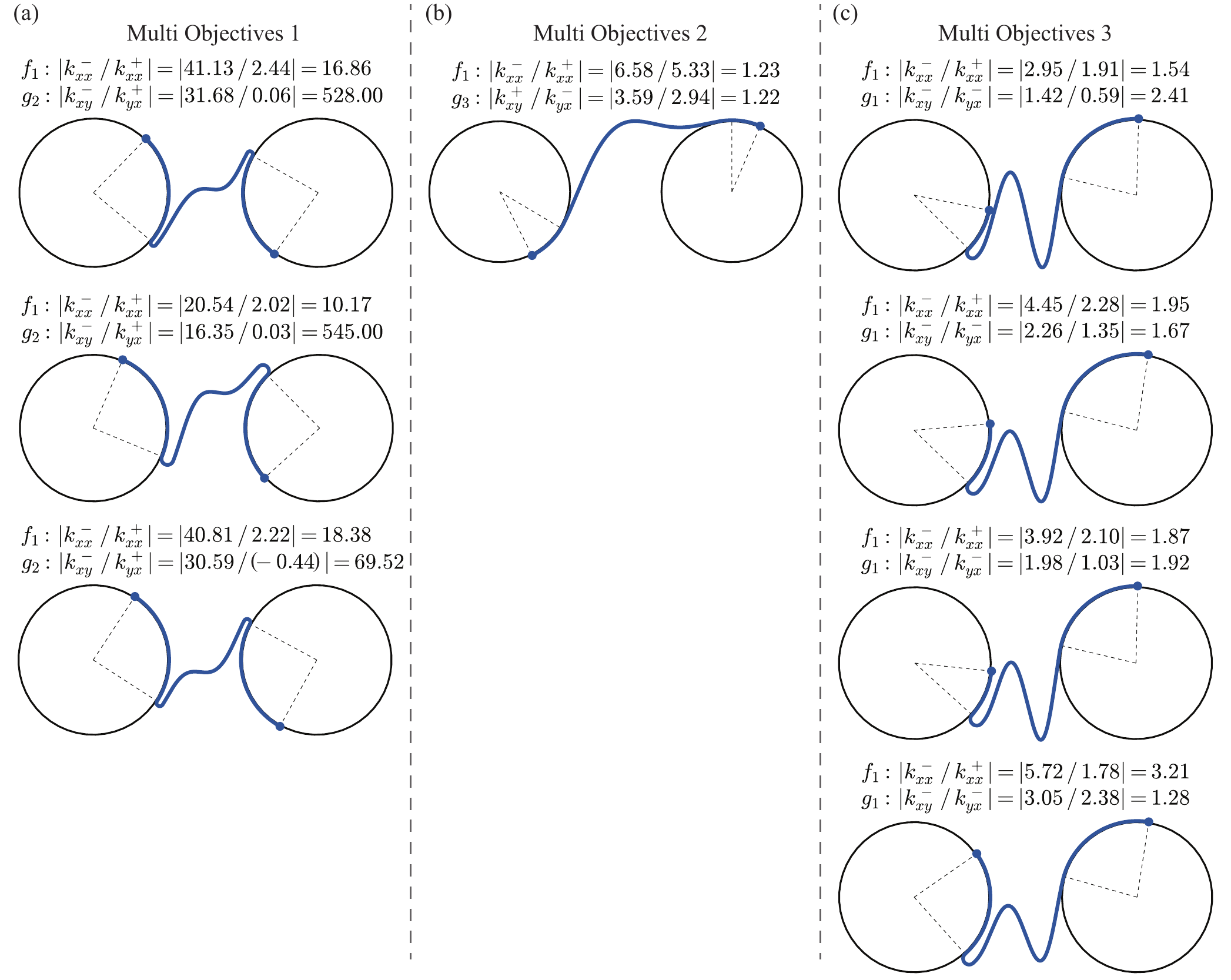}
    \caption{Optimal designs for the three multi-objectives outlined In Section \ref{sec:res_multi}. (a) Three Pareto optimal designs for Multi-Objective 1, discovered from design space 2. (b) One Pareto optimal design for Multi-Objective 2, discovered from design space 1. (c) Four Pareto optimal designs for Multi-Objective 3, discovered from design space 4.}
    \label{fig:apx_multi}
\end{figure}

In Section \ref{sec:res_multi}, we investigated three Multi-Objectives and discussed about the Pareto front for each. Although there can be more than one Pareto Optimum for for each multi-objective, we specifically analyzed the optimal design with largest modulus. Here we present all the Pareto optima for multi-objectives optimization in Fig. \ref{fig:apx_multi}, from which it is evident that the optimal designs for the same multi-objectives share common features, i.e., the shape of the ligament and the connecting area of the circles and ligament appear similar.

\FloatBarrier
\newpage 
\bibliography{references}  

\end{document}